\documentclass[preprint2]{aastex631}
\usepackage{amsmath,amstext}
\usepackage[T1]{fontenc}
\usepackage[figure,figure*]{hypcap}
\usepackage{multirow}
\usepackage[T1]{fontenc} 
\usepackage[utf8]{inputenc} 

\newcommand{\UA}{Steward Observatory, The University of Arizona, 933 N.\ Cherry Ave, Tucson, AZ 85721, USA}
\newcommand{\PSUAA}{Department of Astronomy \& Astrophysics, The Pennsylvania State University, 525 Davey Laboratory, University Park, PA 16802, USA}
\newcommand{\PSUCEHW}{Center for Exoplanets and Habitable Worlds, The Pennsylvania State University, 525 Davey Laboratory, University Park, PA 16802, USA}
\newcommand{\Princeton}{Department of Astrophysical Sciences, Princeton University, 4 Ivy Lane, Princeton, NJ 08540, USA}
\newcommand{\Macquarie}{Department of Physics and Astronomy, Macquarie University, Balaclava Road, North Ryde, NSW 2109, Australia}
\newcommand{\UCI}{Department of Physics \& Astronomy, The University of California, Irvine, Irvine, CA 92697, USA}

\newcommand{\NISTAssoc}{Associate of the National Institute of Standards and Technology, 325 Broadway, Boulder, CO 80305, USA}
\newcommand{\CUBoulder}{Department of Physics, University of Colorado, 2000 Colorado Avenue, Boulder, CO 80309, USA}
\newcommand{\UT}{McDonald Observatory and Center for Planetary Systems Habitability, The University of Texas at Austin, Austin, TX 78730, USA}

\newcommand{\unit}[1]{\ensuremath{\, \mathrm{#1}}}

\shortauthors{Ca\~nas et al.}
\shorttitle{An eccentric, eclipsing MD-BD system}

\begin{document}
\title{An eccentric Brown Dwarf eclipsing an M dwarf}
\correspondingauthor{Caleb I. Ca\~nas}
\email{canas@psu.edu}

\author[0000-0003-4835-0619]{Caleb I. Ca\~nas}
\altaffiliation{NASA Earth and Space Science Fellow}
\affiliation{\PSUAA}
\affiliation{\PSUCEHW}

\author[0000-0001-9596-7983]{Suvrath Mahadevan}
\affil{\PSUAA}
\affil{\PSUCEHW}

\author[0000-0003-4384-7220]{Chad F.\ Bender}
\affil{\UA}

\author[0000-0002-0289-3135]{Noah Isaac Salazar Rivera}
\affil{\UA}

\author[0000-0002-0048-2586]{Andrew Monson}
\affil{\PSUAA}

\author[0000-0001-7708-2364]{Corey Beard}
\affil{\UCI}

\author[0000-0001-8342-7736]{Jack Lubin}
\affil{\UCI}

\author[0000-0003-0149-9678]{Paul Robertson}
\affil{\UCI}

\author[0000-0002-5463-9980]{Arvind F. Gupta}
\affil{\PSUAA}
\affil{\PSUCEHW}

\author[0000-0001-9662-3496]{William D. Cochran}
\affil{\UT}

\author[0000-0002-0560-1433]{Connor Fredrick}
\affil{\NISTAssoc}
\affil{\CUBoulder}

\author[0000-0002-1664-3102]{Fred Hearty}
\affil{\PSUAA}
\affil{\PSUCEHW}

\author[0000-0002-7227-2334]{Sinclaire Jones}
\affiliation{\Princeton}

\author[0000-0001-8401-4300]{Shubham Kanodia}
\affiliation{\PSUAA}
\affiliation{\PSUCEHW}

\author[0000-0002-9082-6337]{Andrea S.J. Lin}
\affil{\PSUAA}
\affil{\PSUCEHW}

\author[0000-0001-8720-5612]{Joe P. Ninan}
\affiliation{\PSUAA}
\affiliation{\PSUCEHW}

\author[0000-0002-4289-7958]{Lawrence W. Ramsey}
\affil{\PSUAA}
\affil{\PSUCEHW}

\author[0000-0002-4046-987X]{Christian Schwab}
\affil{\Macquarie}

\author[0000-0001-7409-5688]{Gu\dh mundur Stef\'ansson}
\altaffiliation{Henry Norris Russell Fellow}
\affiliation{\Princeton}

\begin{abstract}
We report the discovery of a \(M=67\pm2 \mathrm{M_J}\) brown dwarf transiting the early M dwarf TOI-2119 on an eccentric orbit ($e=0.3362 \pm 0.0005$) at an orbital period of \(7.200861 \pm 0.000005\) days. We confirm the brown dwarf nature of the transiting companion using a combination of ground-based and space-based photometry and high-precision velocimetry from the Habitable-zone Planet Finder. Detection of the secondary eclipse with TESS photometry enables a precise determination of the eccentricity and reveals the brown dwarf has a brightness temperature of \(2100\pm80\) K, a value which is consistent with an early L dwarf. TOI-2119 is one of the most eccentric known brown dwarfs with $P<10$ days, possibly due to the long circularization timescales for an object orbiting an M dwarf. We assess the prospects for determining the obliquity of the host star to probe formation scenarios and the possibility of additional companions in the system using Gaia EDR3 and our radial velocities.

\end{abstract}


\section{Introduction}
Brown dwarfs are objects with radii comparable to Jupiter and masses between \(13\sim80 \mathrm{M_J}\) \citep[see][]{Chabrier2000a,Burrows2001}, although this lower limit is not well defined if these objects are classified on the basis of their formation mechanism \citep[e.g.,][]{Chabrier2014}. As isolated objects, brown dwarfs have traditionally been identified via photometric surveys by their colors \citep[e.g.,][]{Pinfield2008,Zhang2009,Folkes2012,Reyle2018} because their spectral energy distribution peaks in the near-infrared. Brown dwarfs contain complex spectral features that are difficult to model and existing evolutionary models are largely degenerate in age, radius, and metallicity. This makes it difficult to determine fundamental properties, such as the mass and radius, for isolated brown dwarfs. When these objects are eclipsing companions to a main-sequence star, photometric and spectroscopic observations yield a measurement of the mass and radius. 

Brown dwarfs are infrequent companions ($\lesssim 1\%$) to main sequence stars \citep[e.g.,][]{Vogt2002,Patel2007,Wittenmyer2009,Sahlmann2011,Nielsen2019}, but previous radial velocity (RV) surveys have facilitated their study \citep[e.g.,][]{Campbell1988,Marcy2000,Wittenmyer2009,Sahlmann2011,Bonfils2013}. These surveys have revealed the ``brown dwarf desert'' \citep[e.g.,][]{Marcy2000,Grether2006} or the apparent paucity of brown dwarf companions to main sequence stars within \(\sim3\) au. This feature has been attributed to the different formation mechanisms between low and high-mass companion brown dwarfs \citep[see][]{Ma2014,Chabrier2014} where high-mass brown dwarfs (\(\gtrsim43 \mathrm{M_J}\)) are believed to form through molecular cloud fragmentation, similar to a binary stellar companion, while low-mass brown dwarfs form through gravitational instabilities in the protoplanetary disk. 

Another factor that may sculpt the brown dwarf desert is the orbital migration and tidal in-spiral \citep[e.g.,][]{Armitage2002,Paetzold2002,Damiani2016} of brown dwarf companions. \cite{Damiani2016} note that tidal interactions with the host star and angular momentum loss through magnetic braking can lead to rapid in-spiral of a brown dwarf companion orbiting main sequence dwarfs with outer convective envelopes. The in-spiral time scale is a strong function of stellar radius, such that early M dwarfs like TOI-2119, have larger in-spiral timescales than later main sequence dwarfs. This enables close brown dwarf companions to exist for a longer time despite the more efficient tidal dissipation and magnetic braking mechanisms present in M dwarfs. \cite{Carmichael2020} noted that while there was no obvious trend in the frequency of brown dwarfs with stellar host type, a large percent of well-characterized (with mass and radius measurements) brown dwarfs (6/23 or \(\sim26\%\) of the then known brown dwarfs) were found to transit M dwarfs. This appears contrary to the occurrence of Jupiter-sized exoplanets, in which the occurrence rate of these planets decreases for M dwarf host stars \citep[e.g.;][]{Endl2006,Johnson2010,Bonfils2013,Maldonado2020}. Currently, there are 5 well-characterized Jupiter-sized planets transiting M dwarfs and 8 transiting M dwarf - brown dwarf systems \citep[including TOI-2119;][]{Carmichael2020,Artigau2021}. More masses and radii for brown dwarf companions are needed to probe any dependence on stellar host type.

In this paper, we present a new M dwarf system, TOI-2119, hosting a high-mass brown dwarf on a \(\sim7.2\)-day eccentric orbit. We confirm the brown dwarf nature of TOI-2119.01 using space-based photometry from the Transiting Exoplanet Satellite \citep[TESS][]{Ricker2015}, additional ground-based photometry, adaptive optics (AO) imaging with the ShaneAO instrument \citep{Srinath2014} on the 3 m Shane Telescope at Lick Observatory, and near-infrared (NIR) RVs with the northern spectrograph of the APO Galaxy Evolution Experiment \citep[APOGEE-2N;][]{Majewski2017,Zasowski2017} and the Habitable-zone Planet Finder Spectrograph \citep[HPF;][]{Mahadevan2012,Mahadevan2014}. 

The paper is structured as follows: Section \ref{sec:obs} presents the observations used in this paper, Section \ref{sec:spec} describes the method for spectroscopic characterization and our best estimates of the stellar parameters, and Section \ref{sec:tranfit} explains the analysis of the photometric and RV data. A discussion of the bulk properties of TOI-2119 in context of other transiting brown dwarfs with masses is presented in Section \ref{sec:discussion}. We conclude the paper in Section \ref{sec:summary} with a summary of our key results.

\section{Observations}\label{sec:obs}
\subsection{TESS}
TESS observed TOI-2119 (Gaia EDR3 1303675097215915264) in short-cadence mode during Sectors 24 and 25 with data spanning 2020 Apr 16 through 2020 June 08. It has one transiting candidate, TOI-2119.01, identified by the TESS Science Processing Operations Center pipeline \citep[SPOC][]{Jenkins2016} with an orbital period of \(\sim 7.2\) days. The ``quick-look pipeline'' developed by \cite{Huang2020} also detected TOI-2119.01 as a target of interest in the full-frame image data of Sectors 24 and 25. 

\begin{figure*}[!ht]
\epsscale{1.15}
\plotone{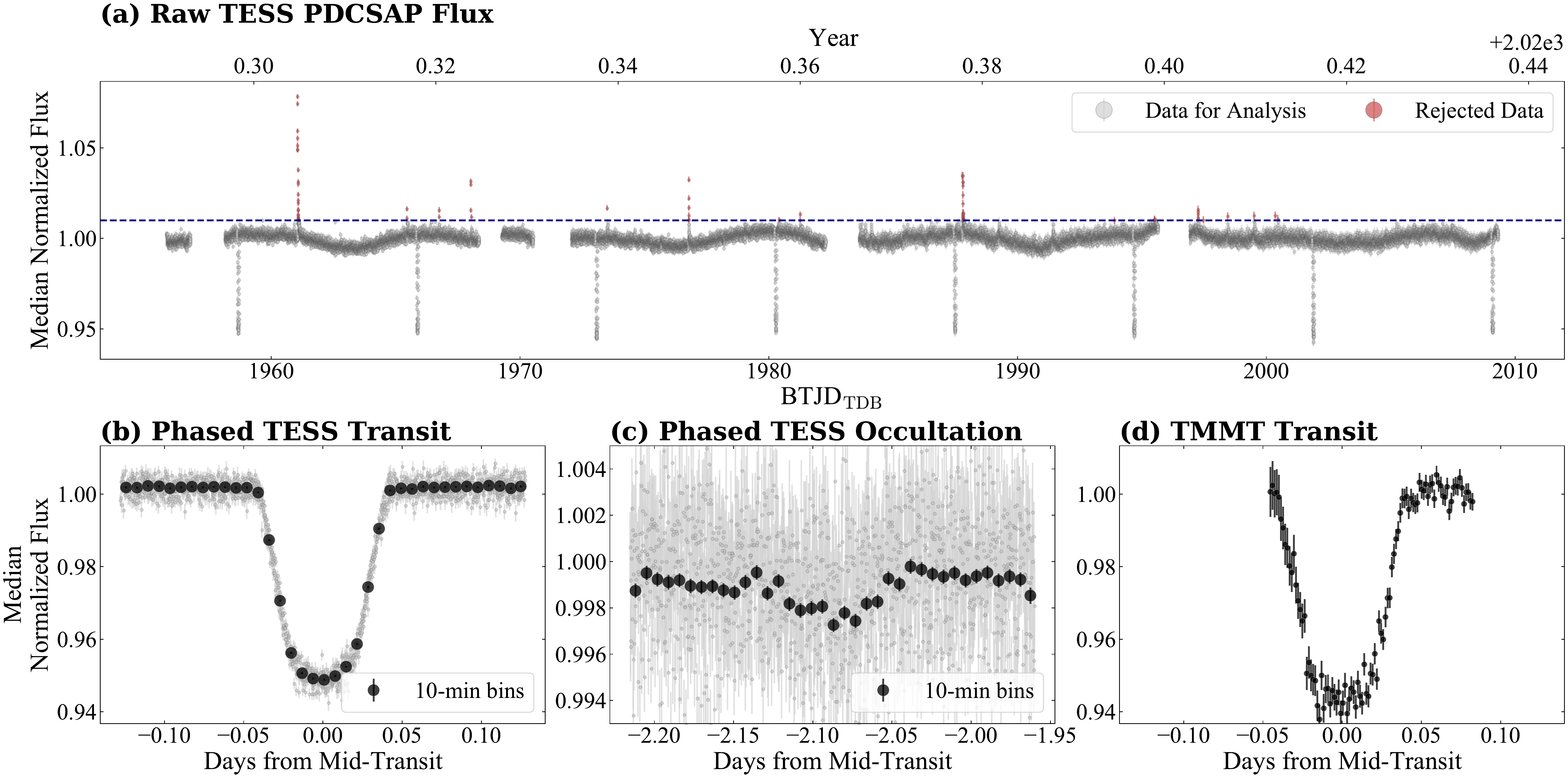}
\caption{\textbf{(a)} displays the PDCSAP photometry from TESS. The light curve reveals a periodic \(\sim5\%\) signal around a flaring, active M dwarf. The dashed line is the threshold of 1.01 which is used to flag data during flares. \textbf{(b)} shows the transit after phasing the raw PDCSAP photometry from (a) to the period and ephemeris determined by the SPOC pipeline. The large circles represent 10-min bins of the phase-folded data. \textbf{(c)} is similar to (b) and presents the TESS occultation. \textbf{(d)} shows the phase-folded TMMT photometry.}
\label{fig:2119phot}
\end{figure*}

For this work, we use the pre-search data-conditioned \citep[PDCSAP;][]{Jenkins2016} light curves available at the Mikulski Archive for Space Telescopes (MAST). The PDCSAP photometry is corrected for instrumental systematics and dilution from other objects contained within the aperture using algorithms that were originally developed for the Kepler mission \citep{Stumpe2012,Smith2012}. Observations with non-zero data quality flags that indicate anomalous data due to various conditions, such as spacecraft events or cosmic ray hits, are excluded from the analysis. The quality flags are described in the TESS Science Data Products Description Document \citep[Table 28 in][]{Tenenbaum2018}. Figure \ref{fig:2119phot} displays the photometry and the transits observed by TESS. The TESS photometry reveals TOI-2119 to be an active, flaring star and, to remove the flares, we reject any median normalized observation larger than 1.01. We perform no additional outlier rejection beyond the data quality flags and application of a threshold value.

\subsection{TMMT}
We observed one transit of TOI-2119.01 using the robotic Three-hundred MilliMeter Telescope \citep[TMMT;][]{Monson2017} at Las Campanas Observatory (LCO) on the night of 2021 April 13. The observations were performed slightly out of focus in the Bessell $I$ filter \citep{Bessell1990} with a point spread function FWHM of $4\arcsec$. We obtained 107 frames using an exposure time of 100s while operating in a 1x1 binning mode. In this mode, TMMT has a 13s readout time between exposures, resulting in an effective cadence of 113s and an observing efficiency of 90\%. The observations began at airmass 5.29 and ended at airmass 1.76.

We processed the photometry using AstroImageJ \citep{Collins2017} following the procedures described in \cite{Stefansson2017}. The light curve was extracted using simple aperture photometry with an object aperture radius of 6 pixels (7.2\arcsec), and inner and outer sky annuli of 10 pixels (12\arcsec) and 15 pixels (17.9\arcsec), respectively, because these values minimized the standard deviation in the residuals. Following \cite{Stefansson2017}, we added the expected scintillation-noise errors in quadrature to the photometric error (including photon, readout, dark, sky background, and digitization noise). Figure \hyperref[fig:2119phot]{1(d)} displays the photometry from TMMT.

\subsection{Doppler Spectroscopy with APOGEE-2N}
TOI-2119 was observed from the Apache Point Observatory on 11 and 12 May 2018 using APOGEE-2N, a multiplexed, high-resolution (\(R\sim 22,500\), NIR (\(\lambda \sim 1.5-1.7\) micron) fiber-fed spectrograph \citep{Wilson2012,Wilson2019} that is mounted on the Sloan 2.5-meter telescope \citep{Gunn2006}. The APOGEE-2N spectrograph was used as part of a survey in SDSS-IV \citep{Blanton2017} with the primary goal of studying the galactic evolution of the Milky Way through the chemical and dynamical analysis of various stellar populations and Galactic regions. 

For this work, we use the publicly available DR16 \citep{Joensson2020} data of TOI-2119. The APOGEE data pipeline \citep{Nidever2015} performs sky subtraction, telluric and barycentric correction, and wavelength and flux calibration for each observation of TOI-2119. The RVs were derived following the procedure described in \cite{Canas2019}. Briefly, we identified the best-fit synthetic spectrum by cross-correlating the highest S/N spectra using synthetic spectra generated from MARCs models \citep{Gustafsson2008} that were specifically generated for the APOGEE-2N survey \citep[see][]{Meszaros2012,Zamora2015,Holtzman2018}. The synthetic spectrum with the largest correlation was used in the final cross-correlation to obtain the reported radial velocities in Table \ref{tab:toirvs}. The uncertainties for each observation were calculated by following the maximum-likelihood approach presented by \cite{Zucker2003}. The derived RVs, the \(1\sigma\) uncertainties, and the S/N per resolution element ($\sim 2$ pixels) are presented in Table \ref{tab:toirvs}.

\subsection{High-resolution Doppler Spectroscopy with HPF}
We obtained eight 945-second visits of TOI-2119 using HPF with a median signal-to-noise ratio (S/N) per 1D extracted pixel of 168 at $1000\unit{nm}$. HPF is a high-resolution ($R\sim55,000$), fiber-fed \citep{Kanodia2018a}, NIR (\(\lambda\sim8080-12780\)\ \AA) spectrograph located on the 10m Hobby-Eberly Telescope (HET) at McDonald Observatory in Texas \citep{Mahadevan2012,Mahadevan2014} that achieves a long-term temperature stability of $\sim$1$\unit{mK}$ \citep{stefansson2016}. The observations span 19 September 2020 through 26 May 2021 and were executed in a queue by the HET resident astronomers \citep{Shetrone2007}. 

The \texttt{HxRGproc} tool was used to process the raw HPF data and perform bias noise removal, nonlinearity correction, cosmic-ray correction, and slope/flux and variance image calculation \citep{Ninan2018}. The one-dimensional spectra were reduced using the procedures in \cite{Ninan2018}, \cite{Kaplan2019}, and \cite{Metcalf2019}. The wavelength solution and drift correction applied to the data was extrapolated from laser frequency comb (LFC) frames that were taken as part of standard evening and morning calibrations and from LFC calibration frames obtained periodically throughout the night. The extrapolation from LFC frames enables precise wavelength calibration on the order of $<30 \unit{cm/s}$ \citep{Stefansson2020}, a value which is smaller than the photon noise for TOI-2119 (median RV uncertainty of \(48.5\) m/s). 

We used a modified version \citep[see][]{Stefansson2020} of the \texttt{SERVAL} code \citep[SpEctrum Radial Velocity AnaLyser;][]{Zechmeister2018} to calculate RVs. \texttt{SERVAL} employs the template-matching technique to derive RVs \citep[e.g.,][]{Anglada-Escude2012} and creates a master template from the observations to determine the Doppler shift by minimizing the \(\chi^2\) statistic. The master template was generated from all observed spectra while masking telluric regions identified using a synthetic telluric-line mask generated from \texttt{telfit} \citep{Gullikson2014}. The barycentric correction for each epoch was calculated using \texttt{barycorrpy} \citep{Kanodia2018} which implements the algorithms from \cite{Wright2014}. The derived HPF RVs, the \(1\sigma\) uncertainties, and the S/N per pixel at 1000 nm for TOI-2119 are presented in Table \ref{tab:toirvs}.

\begin{deluxetable}{lrcc}
\tablecaption{RVs of TOI-2119. \label{tab:toirvs}}
\tablehead{
\colhead{$\unit{BJD_{TDB}}$}  &
\colhead{RV} &
\colhead{$\sigma$} & 
\colhead{S/N} \\
\colhead{}  & 
\colhead{$(\unit{m/s})$} & 
\colhead{$(\unit{m/s})$} &
\colhead{}
}
\startdata
\multicolumn{4}{l}{\hspace{-0.2cm} APOGEE-2N:}  \\
~~2458249.80011 &  $-18688$ &  213 &  226\tablenotemark{a} \\
~~2458250.84530 &  $-11816$ &  209 &  163 \\
\multicolumn{4}{l}{\hspace{-0.2cm} HPF:}  \\
~~2459111.58596 &  $-17407$ &  15 & 108\tablenotemark{b} \\
~~2459117.57945 &  $-12063$ &  17 &  93 \\
~~2459231.02335 &    1908 &  10 & 164 \\
~~2459232.02920 &   $-4353$ &   9 & 175 \\
~~2459238.00846 &    $-319$ &   9 & 169 \\
~~2459247.00784 &  $-10639$ &   8 & 197 \\
~~2459300.84858 &  $-16666$ &  13 & 124 \\
~~2459360.90067 &    2871 &   7 & 226 \\
\enddata
\tablenotetext{a}{The APOGEE-2N S/N is the median value per resolution element (\(\sim2\) pixels). For comparison, HPF has a resolution \(2.5-3\) times that of APOGEE-2N with a resolution element and PSF that each span \(\sim3\) pixels.}
\tablenotetext{b}{The HPF S/N is the median value per 1D extracted pixel at 1000 nm. All exposure times for HPF are 945 s.}
\end{deluxetable}

\subsection{Adaptive Optics Imaging}
\begin{figure*}[!ht]
\epsscale{1.15}
\plotone{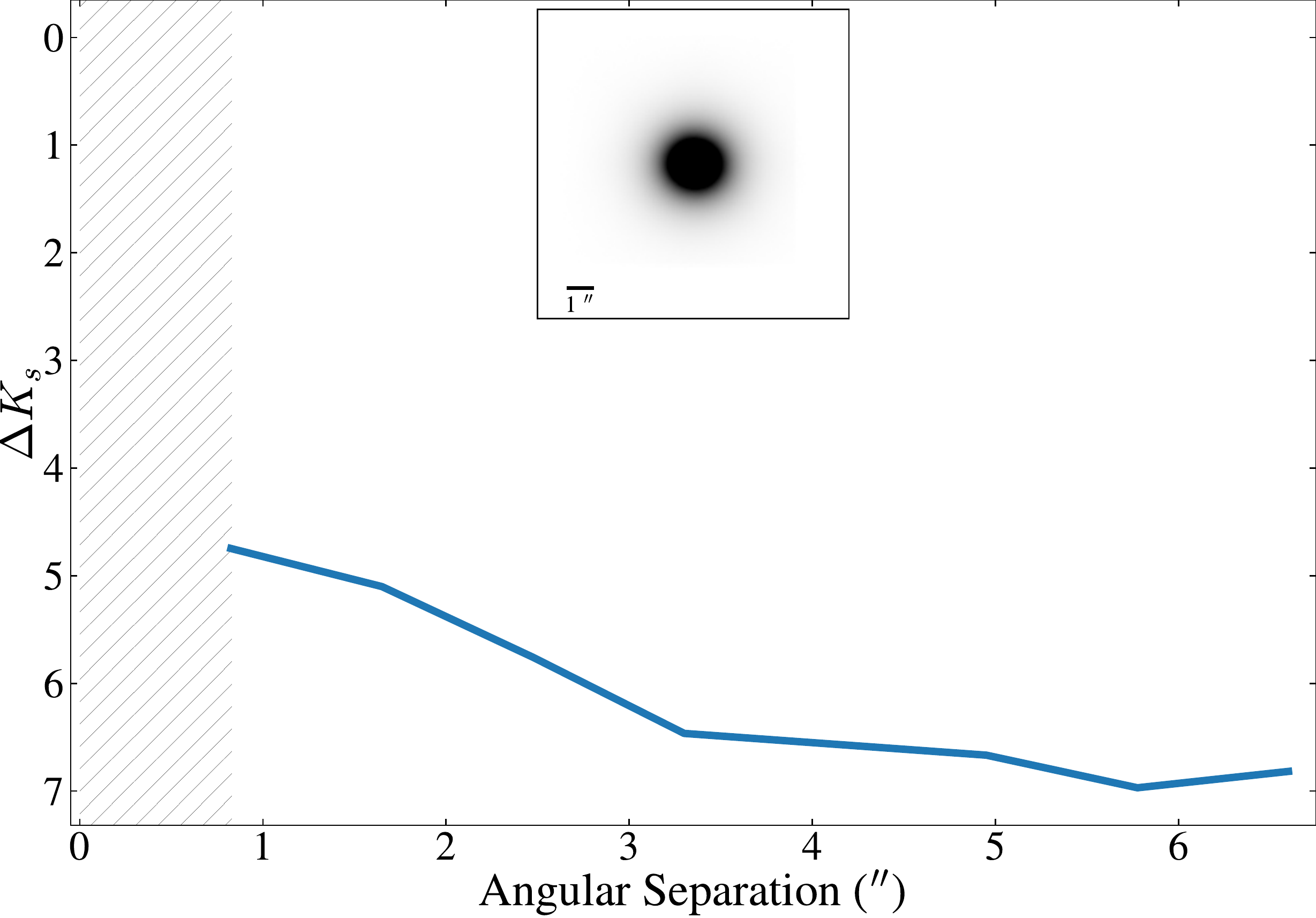}
\caption{Above is the \(5\sigma\) contrast curve obtained using the ShARCS camera in the $K_s$ filter. ShaneAO cannot place constraints within the hatched region ($<0.83\arcsec{}$) due to poor seeing ($\sim2\arcsec{}$) during the observation. The data show there are no bright companions ($\Delta K_s < 4$) at separations of $0.83\arcsec{}-6.5\arcsec{}$ from the host star. The inset image is a cutout centered on TOI-2119 where the scale bar reflects 1\arcsec .}
\label{fig:2119ao}
\end{figure*}

TOI-2119 was observed in the $K_s$ band with the ShARCS camera on the Shane 3m telescope at Lick Observatory \citep{Srinath2014}. It was observed under natural guide star mode using the 5-point dither process described in \cite{Furlan2017}. The seeing was \(\sim2\arcsec{}\) during the observations. The data were reduced using a custom AO pipeline developed that rejects all overexposed or underexposed images and excludes data that are flagged as erroneous (e.g., due to lost guiding, shutters closed early due to weather, etc.). The pipeline applies standard dark correction, flat correction, and a sigma clipping process to all images. A master sky image is produced from the 5-point dither process which is subtracted from each image. The final image is produced by interpolating all images onto a single centroid. A \(5\sigma\) contrast curve is generated from the final image using the algorithms described in \cite{Espinoza2016} and is presented in Figure \ref{fig:2119ao}. The poor seeing during the night TOI-2119 was observed prevents any constraints $<0.83\arcsec{}$ (marked as the hatched region in Figure \ref{fig:2119ao}). There are no bright companions ($\Delta K_s < 4$) that could be a source of contamination in the photometry at separations of $0.83\arcsec{}-6.5 \arcsec{}$ from TOI-2119.

\subsection{Speckle Imaging}
\begin{figure*}[!ht]
\epsscale{1.15}
\plotone{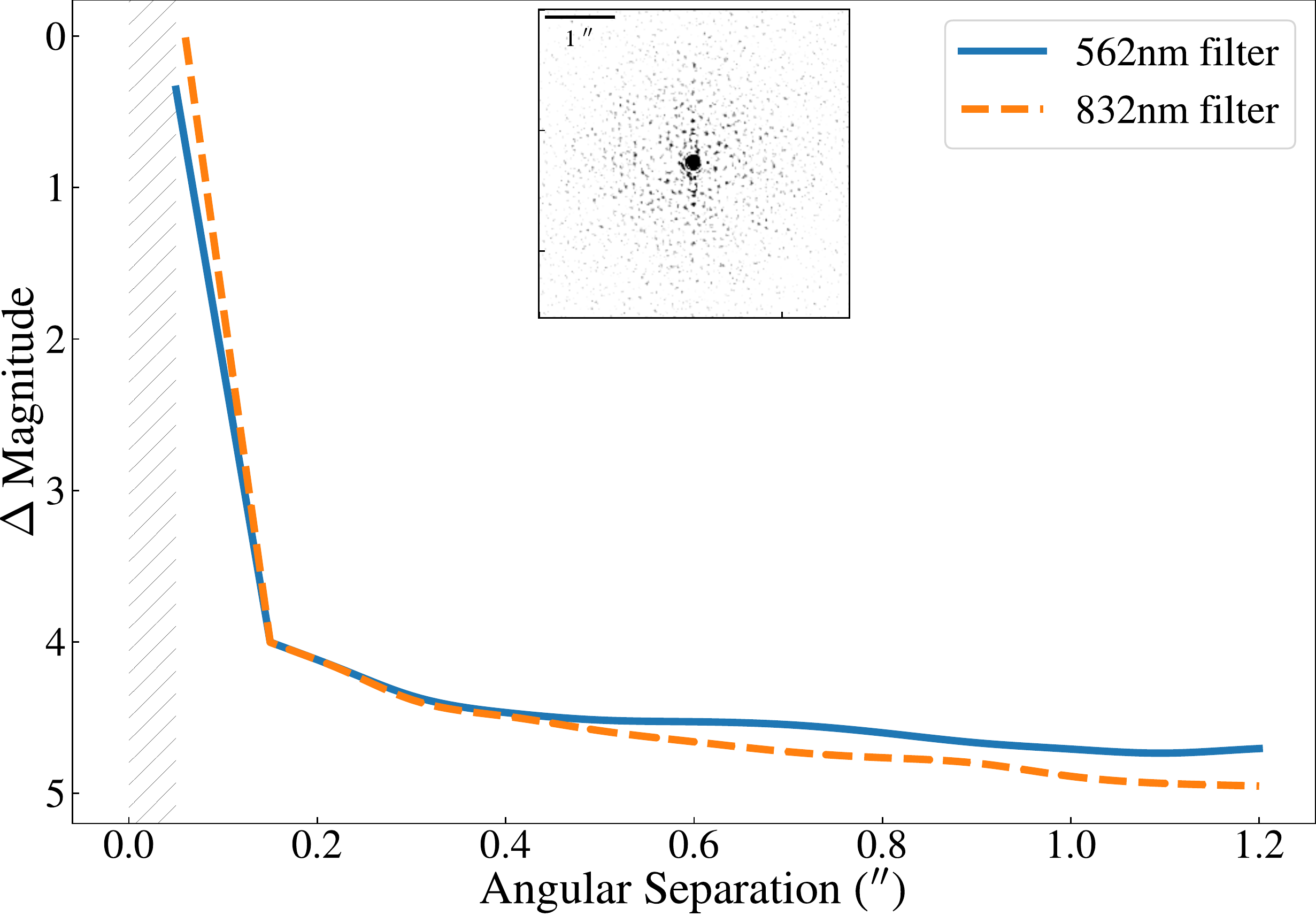}
\caption{The \(5\sigma\) contrast curve obtained using NESSI in the narrowband filters centered at 562nm and 832nm showing no bright companions ($\Delta \mathrm{Mag} < 4$) between \(0.15-1.2''\) from the host star. NESSI cannot place constraints within the hatched region ($<0.05\arcsec{}$) The inset image is the 832nm narrowband filter centered on TOI-2119 where the scale bar reflects 1\arcsec.}
\label{fig:2119speckle}
\end{figure*}

We used NESSI \citep{Scott2018} on the 3.5 m WIYN Telescope at KPNO to perform speckle imaging on 2021 October 26. TOI-2119 was observed in two narrowband filters centered at 562nm and  832nm. The images in each filter were reconstructed following the procedures outlined in \cite{Howell2011}. The NESSI contrast curves in both filters are shown in Figure \ref{fig:2119speckle}, along with an inset of the image in the 832nm narrowband filter. The NESSI data show no evidence of blending from a bright companion $\Delta\mathrm{Mag}<4$ at separations of $0.15-1.2\arcsec$. 

\subsection{Non-detection of Spectroscopic Companions within 2''}
The fibers for APOGEE-2N have a field-of-view of 2\arcsec~ \citep[see][]{Wilson2019} and we use the $H$-band APOGEE-2N spectra to search for light from secondary stars around TOI-2119. We employ the software \texttt{binspec\_plus}, which is based on \texttt{binspec} \citep{El-Badry2018,El-Badry2018a}, to search for the faint spectrum of a second star by modeling the observed spectrum as the sum of two model spectra. \texttt{binspec} was designed to fit both single and double-lined spectra, but its sensitivity is limited to the detection of moderate-mass ratio binaries (\(0.4\lesssim q\lesssim0.85\)) because of the limited temperature regime of the spectral models. \texttt{binspec\_plus}\footnote{\url{https://github.com/tingyuansen/binspec_plus/}} extends the spectral model to include redder dwarfs and giants. We fit the APOGEE-2N spectrum of the TOI-2119 with a neural network spectral model \citep[see][]{Ting2019} trained on the Kurucz stellar library \citep{Kurucz1979} that is valid for slowly rotating (\(v\sin i <45\unit{km\ s^{-1}}\)) main-sequence stars in the regime of $3000 \mathrm{K} < T_{e} < 7000\unit{K}$, $4.0 < \log g < 5.0$, and $-1 < \mathrm{[Fe/H]} < 0.5$. The model selection criterion from Table B1 of \cite{El-Badry2018} relies on two values, $\Delta \chi^2 = \chi^2_{\mathrm{single}}-\chi^2_{\mathrm{binary}}$, which quantifies how better a fit is obtained by the binary model, and the improvement fraction, $f_{imp}$, which describes how better the binary model fit is relative to how different it is from a single-star model. TOI-2119 is classified as a single-lined spectroscopic binary (SB1) because both $\Delta\chi^2$ and $f_{imp}$ are improved relative to a single-star model, but a binary component fit is disfavored and shows negligible improvement in the fit when compared to the SB1 fit (\(\Delta\chi^{2}<3000\) and \(f_{imp}<0\)). This analysis of the APOGEE-2N data reveals no evidence for secondary light within 2\arcsec{} from companion or background dwarf stars with $3000 \mathrm{K} < T_{e} < 7000\unit{K}$.

\section{Stellar Parameters}\label{sec:spec}
\subsection{Spectroscopic Parameters}
The spectroscopic stellar parameters ($T_e$, $\log g$, and [Fe/H]) for TOI-2119 were calculated using the \texttt{HPF-SpecMatch}\footnote{\url{https://gummiks.github.io/hpfspecmatch/}} package \cite[][S. Jones et. al. 2022, in prep.]{Stefansson2020}, which employs the empirical template matching methodology discussed in \cite{Yee2017}. \texttt{HPF-SpecMatch} derives the stellar properties by comparing the highest S/N observed spectra to a library of high quality (\(\unit{S/N}>100\)) HPF stellar spectra with well-determined properties \citep[values adopted from][]{Yee2017}. It identifies the best-matching library spectrum using \(\chi^{2}\) minimization, creates a composite spectrum from a weighted, linear combination of the five best-matching library spectra, and derives the stellar properties using these weights. The reported uncertainty for each stellar parameter is the standard deviation of the residuals from a leave-one-out cross-validation procedure applied to the entire HPF library in the chosen spectral order.

The library contains 166 stars and spans the following parameter space: $2700 \unit{K} < T_{e} < 6000 \unit{K}$, $4.3<\log g < 5.4$, and $-0.5 < \mathrm{[Fe/H]} < 0.5$. We used HPF order index 17 (spanning $10460-10570$ \AA) for the spectral matching of TOI-2119 because it has little to no telluric contamination. Table \ref{tab:stellarparam} lists the derived spectroscopic parameters with their uncertainties.

\startlongtable
\begin{deluxetable*}{lccc}
\tablecaption{Summary of Stellar Parameters. \label{tab:stellarparam}}
\tablehead{\colhead{~~~Parameter}&  \colhead{Description}&
\colhead{Value}&
\colhead{Reference}}
\startdata
\multicolumn{4}{l}{\hspace{-0.2cm} Main identifiers:}  \\
~~~TIC &  \(\cdots\)  & 236387002 & TIC \\
~~~Gaia EDR3 & \(\cdots\) & 1303675097215915264 & Gaia EDR3 \\
\multicolumn{4}{l}{\hspace{-0.2cm} Equatorial Coordinates, Proper Motion, Distance, and Maximum Extinction:} \\
~~~$\alpha_{\mathrm{J2016}}$ &  Right Ascension (RA) & 16:17:43.17 & Gaia EDR3 \\
~~~$\delta_{\mathrm{J2016}}$ &  Declination (Dec) & 26:18:15.16 & Gaia EDR3 \\
~~~$\mu_{\alpha}$ &  Proper motion (RA, \unit{mas/yr}) & $-29.27\pm0.02$ & Gaia EDR3 \\
~~~$\mu_{\delta}$ &  Proper motion (Dec, \unit{mas/yr}) & $6.86\pm0.03$ & Gaia EDR3  \\
~~~$d$ &  Distance in pc$^a$  & $31.46\pm0.03$ & Bailer-Jones\\
~~~\(A_{V,max}\) & Maximum visual extinction & $0.01$ & Green\\
\multicolumn{4}{l}{\hspace{-0.2cm} Optical and near-infrared magnitudes:}  \\
~~~$B$ & Johnson $B$ mag & $13.86\pm0.03$ & APASS\\
~~~$V$ & Johnson $V$ mag & $12.37\pm0.04$ & APASS\\
~~~$g'$ & Sloan $g'$ mag & $13.07\pm0.02$ & APASS\\
~~~$r'$ & Sloan $r'$ mag & $11.76\pm0.02$ & APASS\\
~~~$i'$ & Sloan $i'$ mag & $10.75\pm0.02$ & APASS\\
~~~$J$ & $J$ mag & $8.98\pm0.02$ & 2MASS\\
~~~$H$ & $H$ mag & $8.39\pm0.03$ & 2MASS\\
~~~$K_s$ & $K_s$ mag & $8.14\pm0.02$ & 2MASS\\
~~~$W1$ & WISE1 mag & $8.05\pm0.02$ & WISE\\
~~~$W2$ &  WISE2 mag & $7.97\pm0.02$ & WISE\\
~~~$W3$ &  WISE3 mag & $7.88\pm0.02$ & WISE\\
~~~$W4$ &  WISE4 mag & $7.8\pm0.2$ & WISE\\
\multicolumn{4}{l}{\hspace{-0.2cm} Spectroscopic Parameters$^b$:}\\
~~~$T_{e}$ &  Effective temperature in \unit{K} & $3553 \pm 67$& This work\\
~~~$\mathrm{[Fe/H]}$ &  Metallicity in dex & $0.1\pm0.1$ & This work\\
~~~$\log(g)$ & Surface gravity in cgs units & $4.74 \pm 0.04$ & This work\\
\multicolumn{4}{l}{\hspace{-0.2cm} Model-Dependent Stellar SED and Isochrone fit Parameters$^c$:}\\
~~~$M_\star$ &  Mass in $M_{\odot}$ & $0.53\pm0.02$ & This work\\
~~~$R_\star$ &  Radius in $R_{\odot}$ & $0.51\pm0.01$ & This work\\
~~~$\rho_\star$ &  Density in $\unit{g/cm^{3}}$ & $5.7\pm0.4$ & This work\\
~~~$A_v$ & Visual extinction in mag & $0.005\pm0.003$ & This work\\
\multicolumn{4}{l}{\hspace{-0.2cm} Other Stellar Parameters:}           \\
~~~$P_{\mathrm{rot}}$ & Rotation period in days & $13.2 \pm 0.2$ & This work\\
~~~$v\sin i_{\star}$ & Rotational broadening in km/s & $<2$ & This work\\
~~~Age & Age in Gyrs & $0.7-5.1$ & This work\\
~~~RV & Systemic radial velocity in km/s & $-15.72\pm0.02$ & This work \\
\enddata
\tablenotetext{}{References are: TIC \citep{Stassun2019}, Gaia EDR3 \citep{GaiaCollaboration2021}, Bailer-Jones \citep{Bailer-Jones2021}, Green \citep{Green2019}, APASS \citep{Henden2018}, 2MASS \citep{Cutri2003}, WISE \citep{Wright2010}}
\tablenotetext{a}{Geometric distance from \cite{Bailer-Jones2021}.}
\tablenotetext{b}{Derived using the \texttt{HPF-SpecMatch} algorithm.}
\tablenotetext{c}{\texttt{EXOFASTv2} derived values using MIST isochrones.}
\end{deluxetable*}

\subsection{Spectral Energy Distribution Fitting}
We modeled the spectral energy distribution (SED) to derive model-dependent stellar parameters using the {\tt EXOFASTv2} analysis package \citep{Eastman2019}. Our analysis uses the MIST model grid \citep{Dotter2016,Choi2016} which is based on the ATLAS12/SYNTHE stellar atmospheres \citep{Kurucz1970,Kurucz1993}. {\tt EXOFASTv2} calculates the bolometric corrections for the SED fit by linearly interpolating the precomputed bolometric corrections provided by the MIST team in a grid of \(\log g\), \(\mathrm{T_{eff}}\), [Fe/H], and \(A_V\)\footnote{\url{http://waps.cfa.harvard.edu/MIST/model_grids.html\#bolometric}}. 

The fit uses Gaussian priors on the (i) 2MASS \(JHK\) magnitudes, Sloan \(g^\prime, r^\prime, i^\prime\) magnitudes and Johnson \(BV\) magnitudes from \cite{Henden2018}, and Wide-field Infrared Survey Explorer magnitudes \citep{Wright2010}; (ii) host star surface gravity, temperature, and metallicity derived from \texttt{HPF-SpecMatch}; and (iii) the geometric distance calculated from \cite{Bailer-Jones2021}. We apply a uniform prior for the visual extinction in which the upper limit is determined from estimates of Galactic dust \citep{Green2019} calculated at the distance determined by \cite{Bailer-Jones2021}. The \(R_{v}=3.1\) reddening law from \cite{Fitzpatrick1999} is used by \texttt{EXOFASTv2} to convert the extinction determined by \cite{Green2019} to a visual magnitude extinction. Table \ref{tab:stellarparam} contains the stellar priors and derived stellar parameters with their uncertainties. The model-dependent mass and radius for TOI-2119 are \(0.53\pm0.02~\mathrm{M_{\odot}}\) and \(0.51\pm0.01~\mathrm{R_{\odot}}\), respectively.

\subsection{Rotation Period} \label{sec:rotation}
TOI-2119 is a flaring star exhibiting photometric modulations of \(<1\%\) that persist throughout the TESS photometry (see Figure \hyperref[fig:2119phot]{1(a)}). It is chromospherically active as the calcium II infrared triplet lines are observed to be in emission in each HPF observation. We therefore attribute the variability in the light curve to activity-induced photometric modulation and attempt to constrain the rotation period of TOI-2119 using TESS photometry. For this search, we excise all transit events in the TESS data within a window of 0.107 days (1.25 times the transit duration) from the expected mid-transit. We analyzed the TESS data using the generalized Lomb-Scargle periodogram \citep{Zechmeister2009}, the wavelet power spectra \citep[e.g.,][]{Bravo2014}, and the auto-correlation function \citep[e.g.,][]{McQuillan2013,McQuillan2013a}, and estimated the rotation period to be in the range of 5-20 days.

\begin{figure*}[!ht]
\epsscale{1.15}
\plotone{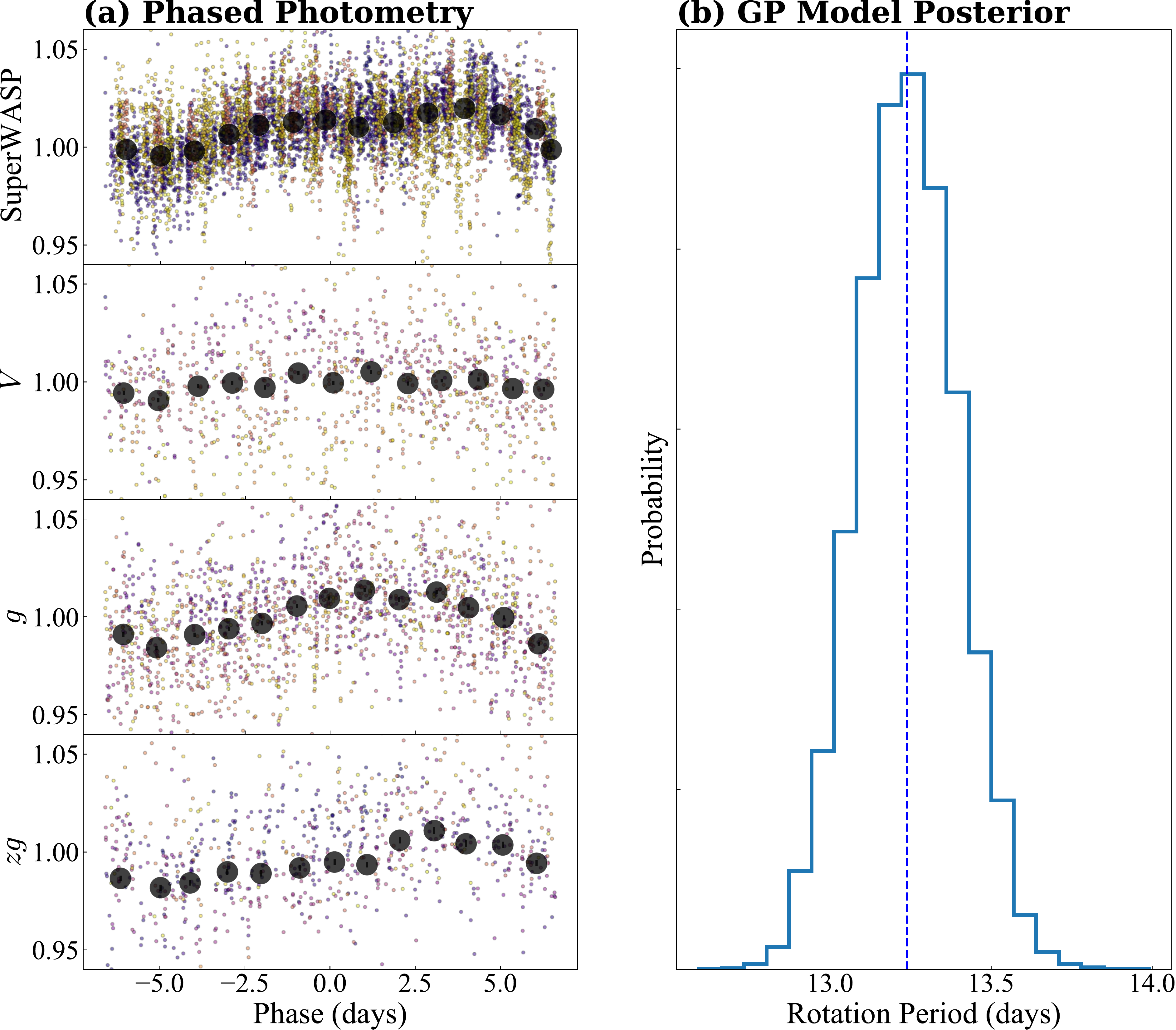}
\caption{\textbf{(a)} is the phased ground-based photometry that was modeled with a Gaussian process. The large black points represent 1-day bins of the phased photometry. \textbf{(b)} presents the posterior distribution of the rotation period of the Gaussian process model. We derive a rotation period of \(13.2\pm0.2\) days.}
\label{fig:2119phasedrot}
\end{figure*}

To further constrain the rotation period, we used publicly available data from the (i) Zwicky Transient Facility \citep[ZTF;][]{Masci2019} in the $zg$ band, (ii) All-Sky Automated Survey for SuperNovae \citep[ASAS-SN;][]{Shappee2014,Kochanek2017} in the $V$ and $g'$ bands, and (iii) SuperWASP \citep{Butters2010}. We modeled the ground-based photometry using the \texttt{juliet} analysis package \citep{Espinoza2019}, which performs the parameter estimation using \texttt{dynesty} \citep{Speagle2020}, a dynamic nested-sampling algorithm. The photometric model is a Gaussian process that uses the approximate quasi-periodic covariance function from the \texttt{celerite} package \citep[Equation 56 in][]{Foreman-Mackey2017} because it has been used to reliably infer stellar rotation rates \citep[e.g.,][]{Angus2018,Robertson2020}. We used our constraints from the TESS photometry to place a uniform prior on the rotation period of $5-20$ days. The fit yields a rotation period of \(13.2\pm0.2\) days and is included in Table \ref{tab:stellarparam}. Figure \ref{fig:2119phasedrot} displays the ground-based photometry used for this analysis and the posterior distribution on the rotation period. 

With a period of \(13.2\pm0.2\) days, TOI-2119 has intermediate rotation period based on the classification scheme of \cite{Newton2016}. \cite{Newton2016} could not provide an age range for M dwarfs with rotation periods spanning 10–70 days but showed that M dwarfs with $P_{\mathrm{rot}}<10$ days have a mean age of $0.7\pm0.3$ Gyr while those with $P_{\mathrm{rot}}>70$ days have mean ages of $5.1^{+4.2}_{-2.6}$ Gyr. TOI-2119 most probably has an age between $0.7-5.1$ Gyr. This is age range is also consistent with the rotation period and age relationship from \cite{Engle2018}.

\section{Photometric and RV Modeling} \label{sec:tranfit}
We use \texttt{allesfitter} \citep{Guenther2021} to jointly model the photometry and RVs. \texttt{allesfitter} calculates the transit and RV models using the \texttt{ellc} package \citep{Maxted2016} and performs the parameter estimation using \texttt{dynesty}. The RV model is a standard Keplerian model while the photometric model is the sum of a transit model, an occultation model, and the same Gaussian process noise model described in Section \ref{sec:rotation} to account for correlated noise in the TESS photometry. The transit model adopts a quadratic limb-darkening law where the limb-darkening coefficients are parameterized following \cite{Kipping2013b} while the occultation model assumes uniform limb darkening. Both the photometric and RV models also include a simple white-noise model in the form of a jitter term that is added in quadrature to the error bars of each instrument.

\begin{figure*}[!ht]
\epsscale{1.15}
\plotone{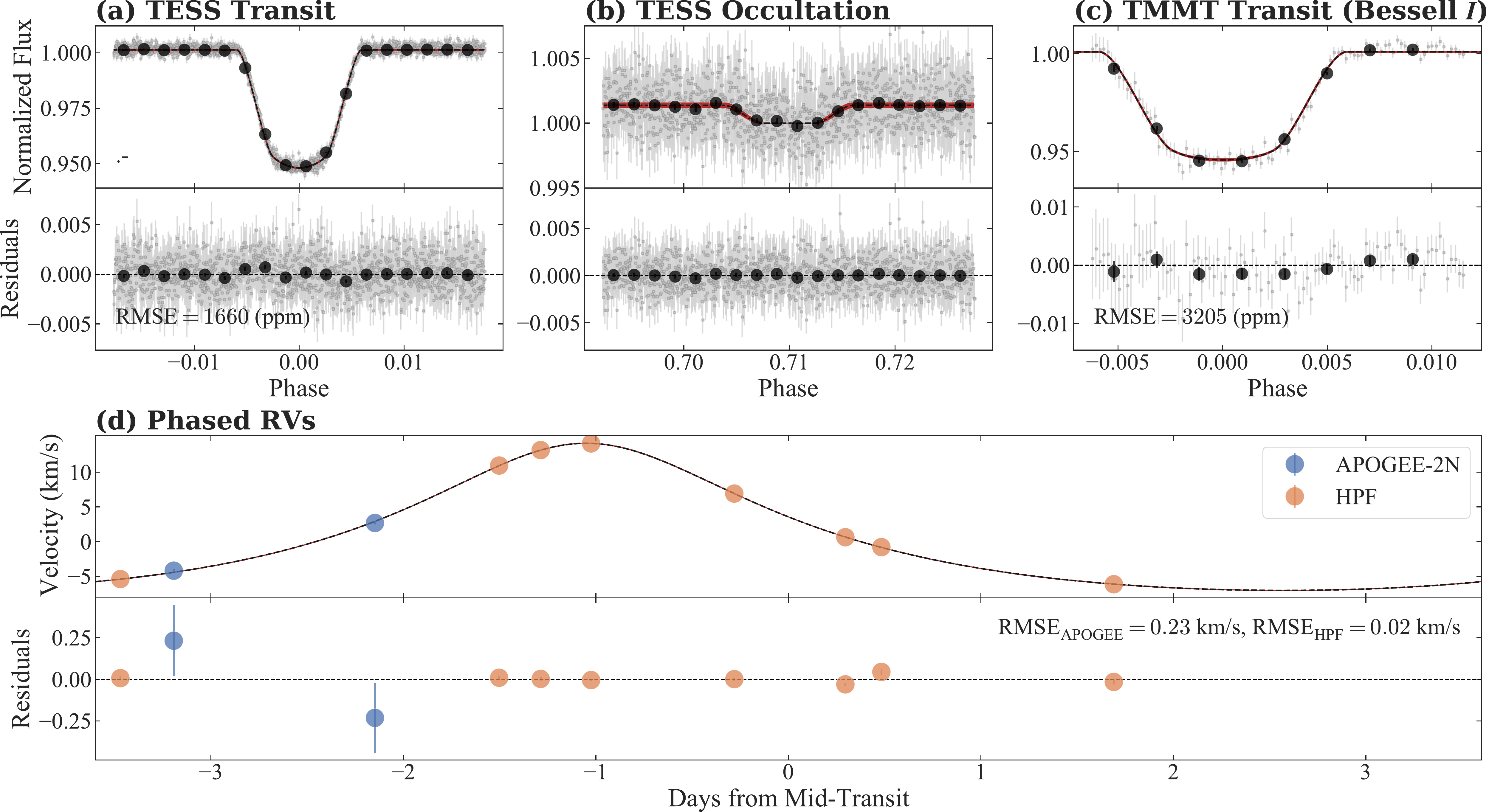}
\caption{Joint Photometry and RV fit for TOI-2119. \textbf{(a)} displays the detrended TESS photometry phased to the derived ephemeris in Table \ref{tab:derivedpar} in the top panel a with the residuals to the best fit transit model in the bottom panel. \textbf{(b)} is identical to (a) but shows the occultation. \textbf{(c)} is similar to (a) and presents the transit fit to the TMMT data. Panels (a)-(c) are plotted in units of orbital phase (from 0-1) where the black points represent 20-min bins of the phase-folded data. \textbf{(d)} shows the phased RVs plotted with the best fit RV model and the corresponding residuals. For all panels, the best-fitting model is plotted as a dashed line while the shaded regions denote the \(1\sigma\) extent of the derived posterior solution.}
\label{fig:rvlcfit}
\end{figure*}

Table \ref{tab:derivedpar} provides a summary of the inferred system parameters and respective confidence intervals and Figure \ref{fig:rvlcfit} displays the model posteriors. The data reveal a high-mass brown dwarf ($M_{2}=67\pm2\mathrm{M_J}$ and $R_{2}=1.11\pm0.03\mathrm{R_J}$) orbiting TOI-2119 on an eccentric orbit ($e=0.3362 \pm 0.0005$) with a period of \(7.200861 \pm 0.000005\) days. The eccentricity of the orbit is determined with exquisite precision because of the powerful constraint on $e$ and $\omega$ from the presence of both a transit and occultation \citep[see][]{Winn2010}.

\startlongtable
\begin{deluxetable*}{llccc}
\tabletypesize{\scriptsize }
\tablecaption{System Parameters for TOI-2119 \label{tab:derivedpar}}
\tablehead{\colhead{~~~Parameter} &
\colhead{Units} &
\colhead{Prior} &
\multicolumn{2}{c}{Value}
}
\startdata
\noalign{\vskip 1.5ex} Photometric Parameters & & & TESS & TMMT\\ \noalign{\vskip .8ex}
~~~Linear Limb-darkening Coefficient$^a$\dotfill & $q_1$\dotfill & $\mathcal{U}(0,1)$ & $0.18_{-0.03}^{+0.05}$ & $0.6 \pm 0.1$ \\
~~~Quadratic Limb-darkening Coefficient$^a$\dotfill & $q_2$\dotfill & $\mathcal{U}(0,1)$ & $0.8_{-0.2}^{+0.1}$ & $0.06_{-0.04}^{+0.06}$ \\
~~~Photometric Jitter\dotfill & $\sigma_{phot}$ (ppm)\dotfill & $\mathcal{J}(10^{-4},10^{6})$ & $0.004^{+0.769}_{-0.004}$ & $10 \pm 10$\\
\noalign{\vskip 1.5ex} RV Parameters & & & APOGEE-2N & HPF\\ \noalign{\vskip .8ex}
~~~RV Offset\dotfill & $\gamma$ (km/s)\dotfill & $\mathcal{U}(-20,-10)$ &
$-14.5\pm0.1$ & $-11.27 \pm 0.01$ \\
~~~RV Jitter\dotfill & $\sigma_{RV}$ (m/s)\dotfill & $\mathcal{J}(10^{-3},10^3)$ & $1 \pm 1$ & $10_{-10}^{+20}$\\
\sidehead{Orbital Parameters:}
~~~Orbital Period\dotfill & $P$ (days) \dotfill & $\mathcal{N}(7.2,0.1)$ & \multicolumn{2}{c}{$7.200861 \pm 0.000005$}\\
~~~Time of Conjunction\dotfill & $T_C$ (BJD\textsubscript{TDB})\dotfill & $\mathcal{N}(2458958.6,0.1)$ & \multicolumn{2}{c}{$2458958.67756 \pm 0.00006$}\\
~~~$\sqrt{e}\cos\omega$\dotfill & \dotfill & $\mathcal{U}(-1,1)$ & \multicolumn{2}{c}{$0.5798 \pm 0.0004$}\\
~~~$\sqrt{e}\sin\omega$\dotfill & \dotfill & $\mathcal{U}(-1,1)$ & \multicolumn{2}{c}{$-0.009 \pm 0.003$}\\
~~~Semi-amplitude velocity\dotfill & $K$ (km/s) \dotfill & $\mathcal{U}(1,20)$ &   \multicolumn{2}{c}{$10.59 \pm 0.02$}\\
~~~Scaled Radius\dotfill & $R_{2}/R_{\star}$ \dotfill & $\mathcal{U}(0,1)$ & \multicolumn{2}{c}{$0.226 \pm 0.001$}\\
~~~$\left(R_\star+R_2\right)/a$\dotfill & \dotfill & $\mathcal{U}(0,1)$ & \multicolumn{2}{c}{$0.0454 \pm 0.0003$}\\
~~~$\cos i$\dotfill & \dotfill & $\mathcal{U}(0,1)$ & \multicolumn{2}{c}{$0.0259 \pm 0.0005$}\\
~~~Surface brightness ratio\dotfill & $J$ \dotfill & $\mathcal{U}(0,1)$ & \multicolumn{2}{c}{$0.027_{+0.004}^{-0.003}$}\\
\sidehead{TESS Gaussian Process Hyperparameters:}
~~~$B$\dotfill & Amplitude ($10^{-6} \unit{ppm}$)\dotfill& $\mathcal{J}(10^{-6},10^6)$ & \multicolumn{2}{c}{$6^{+2}_{-1}$}\\
~~~$C$\dotfill & Additive Factor \dotfill & $\mathcal{J}(10^{-6},10^6)$ & \multicolumn{2}{c}{$90 \pm 90$}\\
~~~$L$\dotfill & Length scale (days) \dotfill & $\mathcal{J}(10^-6,10^6)$ & \multicolumn{2}{c}{$1.5_{-0.3}^{+0.5}$}\\
~~~$P_{GP}$\dotfill & Period (days) \dotfill & $\mathcal{J}(1,100)$ & \multicolumn{2}{c}{$20_{-10}^{+30}$}\\
\sidehead{Derived Parameters:}
~~~Scaled Semi-major Axis\dotfill & $a/R_{\star}$ \dotfill & $\cdots$ & \multicolumn{2}{c}{$27.0 \pm 0.2$}\\
~~~Impact Parameter\dotfill & $b$\dotfill & $\cdots$ & \multicolumn{2}{c}{$0.623 \pm 0.009$}\\
~~~Impact Parameter of occultation\dotfill & $b_s$\dotfill & $\cdots$ & \multicolumn{2}{c}{$0.62 \pm 0.01$}\\
~~~Time of Pericenter\dotfill & $T_p$ (BJD\textsubscript{TDB})\dotfill & $\cdots$ & \multicolumn{2}{c}{$2458963.790 \pm 0.002$}\\
~~~Time of Occultation\dotfill & $T_s$ (BJD\textsubscript{TDB})\dotfill & $\cdots$ & \multicolumn{2}{c}{$2458963.790 \pm 0.002$}\\
~~~Eccentricity\dotfill & $e$ \dotfill & $\cdots$ & \multicolumn{2}{c}{$0.3362 \pm 0.0005$}\\
~~~Argument of Periastron\dotfill & $\omega$ (degrees) \dotfill & $\cdots$& \multicolumn{2}{c}{$-0.9 \pm 0.3$}\\
~~~Orbital Inclination\dotfill & $i$ (degrees)\dotfill & $\cdots$ & \multicolumn{2}{c}{$88.51 \pm 0.03$}\\
~~~Transit Duration\dotfill & $T_{14}$ (hours)\dotfill & $\cdots$ & \multicolumn{2}{c}{$2.039 \pm 0.007$}\\
~~~Occultation Duration\dotfill & $T_{14}$ (hours)\dotfill & $\cdots$ & \multicolumn{2}{c}{$2.025 \pm 0.009$}\\
~~~Mass\dotfill & $M_{2}$  (\unit{M_{J}}) \dotfill & $\cdots$ &  \multicolumn{2}{c}{$67\pm2$}\\
~~~Mass ratio\dotfill & $q=M_{2}/M_{\star}$ \dotfill & $\cdots$ &  \multicolumn{2}{c}{$0.12\pm0.02$}\\
~~~Radius\dotfill & $R_{2}$  (\unit{R_{J}}) \dotfill & $\cdots$ &  \multicolumn{2}{c}{$1.11\pm0.03$}\\
~~~Surface Gravity\dotfill & $\log g_{2}$  (cgs) \dotfill & $\cdots$ &  \multicolumn{2}{c}{$5.158 \pm 0.008$}\\
~~~Density\dotfill & $\rho_{2}$  (g/cm$^3$) \dotfill & $\cdots$ &  \multicolumn{2}{c}{$60\pm5$}\\
~~~Semi-major Axis\dotfill & $a$ (au) \dotfill & $\cdots$ & \multicolumn{2}{c}{$0.064 \pm 0.002$}\\
~~~Transit Depth (TESS bandpass)\dotfill & (\%)\dotfill & $\cdots$ & \multicolumn{2}{c}{$5.09 \pm 0.05$}\\
~~~Occultation Depth (TESS bandpass)\dotfill & ppm ($\times10^{-6}$)\dotfill & $\cdots$ & \multicolumn{2}{c}{$1400\pm200$}\\
~~~$T_{2}$\dotfill & Brightness Temperature (K)\dotfill & $\cdots$ & \multicolumn{2}{c}{$2100\pm80$}\\
\enddata
\tablenotetext{a}{Using the $q1$ and $q2$ parameterization from \cite{Kipping2013b}.}
\end{deluxetable*}

\section{Discussion}\label{sec:discussion}
\subsection{Brightness Temperature}\label{sec:brighttemp}
The depth of the secondary eclipse observed in TESS can be modeled as a function of various fundamental properties \citep[e.g.,][]{Charbonneau2005,Esteves2013,Shporer2017}:
\footnotesize
\begin{equation}
    \mathrm{Depth} = \left(\frac{R_{2}}{R_{\star}}\right)^2 \frac{\int \tau\left(\lambda\right) F_{2,\nu}\left(\lambda,T_{2}\right)d\lambda}{\int \tau\left(\lambda\right) F_{\star,\nu}\left(\lambda,T_{e}\right)d\lambda} + A_g \left(\frac{R_{2}}{a}\right)^2,
    \label{eq:depth}
\end{equation} 
\normalsize
where $\tau\left(\lambda\right)$ is the TESS transmission function\footnote{\url{https://heasarc.gsfc.nasa.gov/docs/tess/the-tess-space-telescope.html\#bandpass}}, $T_{e}$ and $F_{\star,\nu}\left(\lambda,T_{e}\right)$ are the effective temperature and flux of the host star, $T_{2}$ and $F_{2,\nu}\left(\lambda,T_{2}\right)$ is the brightness temperature and flux of TOI-2119.01, and $A_g$ is the geometric albedo. The value of $\left(R_2/a\right)^2=70$ ppm and, even if the geometric albedo were unity, this is a small fraction of the eclipse depth ($1400\pm200$ ppm). \cite{Marley1999} provide a more realistic geometric albedo of $A_g\approx0.1$ for a massive brown dwarf transiting an early M dwarf. This value of $A_g$ would limit the contribution from the second term of Equation \ref{eq:depth} to $\sim7$ ppm. For TOI-2119, we ignore any contribution to the eclipse depth from reflected light because the contribution from these terms are \(\lesssim 10\) ppm, a value which is below the precision of TESS. The contribution from ellipsoidal variations \citep[e.g.,][]{Shporer2017} is similarly negligible and would have an amplitude $\lesssim10$ ppm. To solve for the brightness temperature of TOI-2119.01, we use our posterior distribution of the eclipse depth and estimate the fluxes of TOI-2119 and its brown dwarf companion using the BT-Settl models \citep[][]{Allard2012,Allard2012a} based on the \cite{Caffau2011} solar abundances. To account for the uncertainties in the stellar parameters, we calculate the posterior distribution of the brightness temperature with a Monte Carlo sampling method where we use the stellar parameters derived from our \texttt{HPF-SpecMatch} analysis as Gaussian priors and allow them to vary while deriving the temperature.

This analysis yields a temperature of $T_{2}=2100\pm80$ K for TOI-2119.01. The measured temperature of TOI-2119.01 spans the M-L transition \citep[between 2000-2500 K;][]{Allard2013} and is comparable to those of early L-dwarfs from field \citep[e.g.,][]{Helling2014,Zhang2017} and astrometric discoveries \citep{Dupuy2017}. Future observations of TESS in Sectors 51 and 52 will provide additional eclipses to further improve the eclipse depth and temperature determination. Additional observations of the secondary eclipse in different bandpasses \citep[e.g.,][]{Beatty2014,Croll2015,Beatty2017} are necessary to probe the dayside eclipse spectrum of TOI-2119.01 and compare it to theoretical atmospheric models \citep[e.g.,][]{Beatty2020}.

\subsection{Comparison to the existing brown dwarf population}
\begin{figure*}[!ht]
\epsscale{1.15}
\plotone{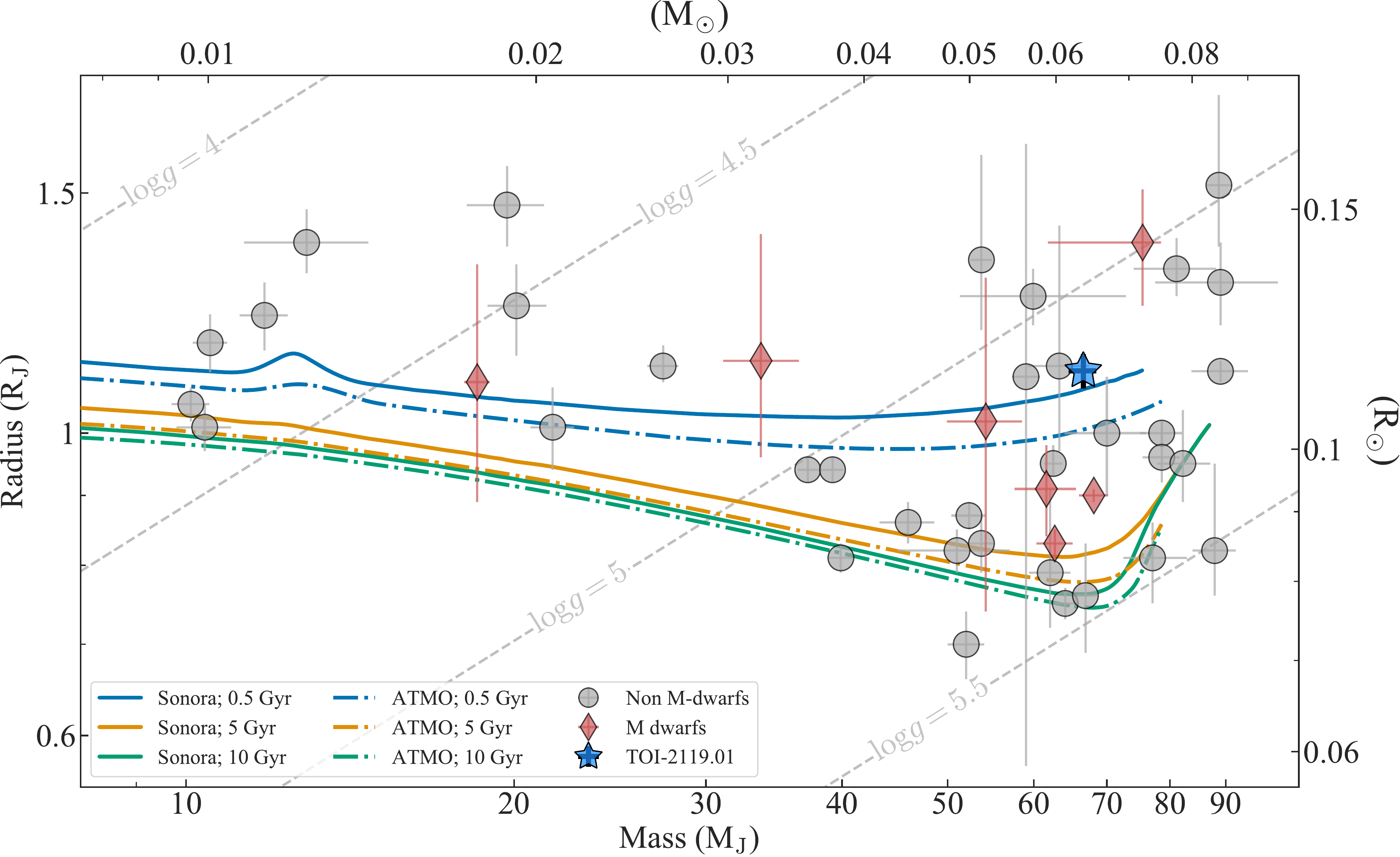}
\caption{The brown dwarf mass-radius diagram showing TOI-2119.01 and all substellar companions from the literature with masses between \(10-90\mathrm{M_J}\) and radii \(<2\mathrm{R_J}\). Grey circles are systems with non-M dwarf hosts while red diamonds are brown dwarfs transiting M dwarfs. TOI-2119 is plotted as the blue star. Brown dwarfs with radii \(>2\mathrm{R_J}\) are found in young clusters with ages \(\ll1\) Gyr. The solid lines (Sonora) are the cloudless, solar metallicity evolutionary tracks from \cite{Marley2021} and the dash-dotted lines (ATMO) are the evolutionary tracks from \cite{Phillips2020} at ages 0.5, 5, and 10 Gyr. Contour lines of fixed \(\log g\) values are included for reference.}
\label{fig:bdmassrad}
\end{figure*}

An analysis of the Kepler \citep{Santerne2016} and CoRoT \citep{Csizmadia2016} samples of short-period transiting brown dwarfs revealed that they are rare with an occurrence rate of $\sim0.2-0.3\%$ around Sun-like stars. Including TOI-2119.01, there are 48 known transiting brown dwarfs, 9 of which have M dwarf host stars. Figure \ref{fig:bdmassrad} shows the location of TOI-2119.01 on a mass-radius diagram of transiting brown dwarfs compiled from the literature \citep{Carmichael2020,Mireles2020,Casewell2020,Grieves2021}. For comparison, we include the cloudless, solar metallicity evolutionary models by \cite{Marley2021} and the evolutionary models by \cite{Phillips2020}, each calculated in chemical equilibrium, at age of 0.5, 5 and 10 Gyr. 

TOI-2119.01 is located in a cluster of other high-mass brown dwarfs and appears to be consistent (within $2\sigma$) with brown dwarf models for ages $<1$ Gyr. This age is within the range of $0.7-5.1$ Gyr determined from the rotation period (see Section \ref{sec:rotation}). The small ($<2\sigma$) discrepancy between the observed mass and radius and the predicted values for models between $0.7-5.1$ Gyrs is also seen with different evolutionary tracks, such as the models from \cite{Baraffe2003} and \cite{Saumon2008}. Figure \ref{fig:bdmassrad} shows that the brown dwarf models evolve quickly for objects $<5$ Gyr, emphasizing the need for precise ages to accurately discriminate between these models. A small discrepancy is not surprising because of the approximate age determination and the complexity of model atmospheres and the equation of state for these ultracool objects \cite[see][]{Burrows2011,Chabrier2019,Phillips2020,Marley2021}. Some of the observed inflation in radii for very low-mass stars and brown dwarfs may be attributed to strong magnetic activity, which would inhibit efficient convection \citep[e.g.,][]{Lopez-Morales2007,Torres2010,Stassun2012,MacDonald2017} and result in a larger radius when compared to evolutionary models. We have also ignored the role of the stellar insolation for the shortest period brown dwarfs. Additional well characterized transiting brown dwarfs, particularly those with known ages, are necessary to further improve evolutionary tracks of brown dwarfs.

\begin{figure*}[!ht]
\epsscale{1.15}
\plotone{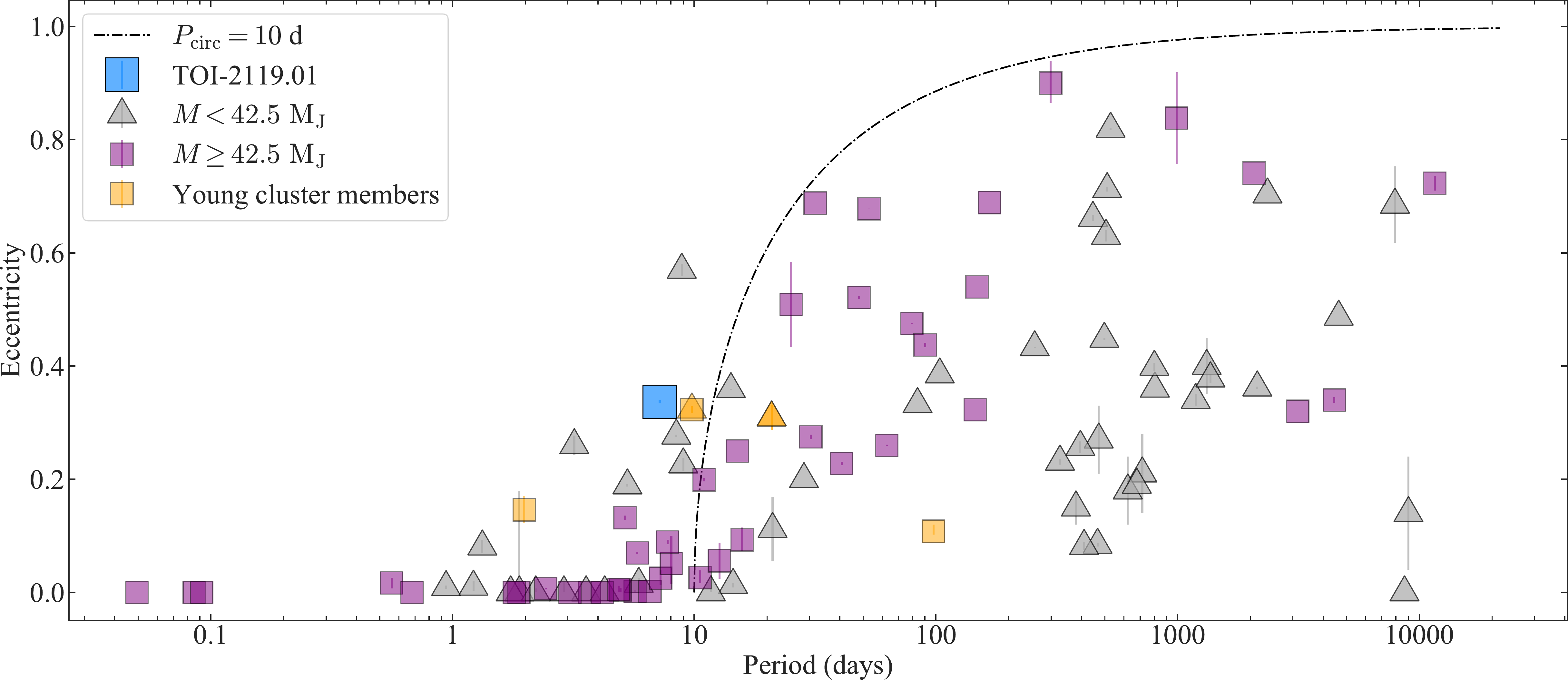}
\caption{The eccentricity as a function of the period for brown dwarfs from the catalogues of \cite{Ma2014}, \cite{Carmichael2020}, \cite{Mireles2020}, \cite{Casewell2020}, and \cite{Grieves2021}. Triangles denote the brown dwarfs with masses smaller than 42.5 $\mathrm{M_J}$ while the squares are larger than this mass. Systems in young clusters: 2M0535-05A/B \citep{Stassun2006}, AD 3116 \citep{Gillen2017}, RIK 72 \citep{David2019}, and 2M1510Aa/b \citep{Triaud2020} are denoted with yellow markers. The dashed line indicates the maximum eccentricity for systems unaffected by tides when adopting a circularization period of 10 days. TOI-2119.01, the blue square, is most eccentric high-mass brown dwarf discovered with a period $<10$ days.}
\label{fig:bdperecc}
\end{figure*}

From a statistical analysis of the brown dwarf population, \cite{Ma2014} postulated the brown dwarf sample is comprised of two populations: (i) a low-mass group with $M<42.5\mathrm{M_J}$ with an eccentricity distribution comparable to gas giants and (ii) a high-mass group with an eccentricity distribution comparable to binary stars. Figure \ref{fig:bdperecc} compares TOI-2119.01 on the period-eccentricity diagrams with the transiting brown dwarfs described above and from the catalogue compiled by \cite{Ma2014}. TOI-2119.01 is the most eccentric high-mass brown dwarf with a period $<10$ days. The period and eccentricity of TOI-2119 is inconsistent with the $\sim10$ day circularization period (plotted as a dashed line in Figure \ref{fig:bdperecc}) observed in M dwarf binaries \citep[e.g.,][]{Udry2000,Mayor2001} and the $\sim10-12$ day circularization period observed in sun-like binaries \citep[e.g.,][]{Duquennoy1991,Meibom2005,Raghavan2010}. We note that two systems in young clusters with ages $<1$Gyr, the 2M0535-05 \citep{Stassun2006} and AD 3116 \citep{Gillen2017} systems, host massive brown dwarfs with non-zero eccentricities with $P<10$ days, but this is not surprising given their young ages. TOI-2119.01 is consistent with the shorter circularization period of \(\sim 3-5\) days that has been suggested for the giant exoplanet population \citep[e.g.,][]{Halbwachs2005,Pont2011,Bonomo2017}. 

\subsection{Potential Formation Mechanisms}
There are various mechanisms of formation for brown dwarfs \citep[see][]{Whitworth2007,Chabrier2014,Whitworth2018} including gravitational instability and turbulent fragmentation of a molecular cloud \citep{Padoan2002,Padoan2004,Hennebelle2008}, disk instability and migration \citep[e.g.,][]{Helled2014,Kratter2016,Nayakshin2017,Mueller2018}, and core accretion for low-mass brown dwarfs \citep[e.g.,][]{Lambrechts2012,Molliere2012}. The mass of TOI-2119.01 means it may have been formed through gravitational instability in a disk, and simulations by \cite{Forgan2018} show that dynamical interactions and scattering between fragments in a gravitationally unstable disk can readily form brown dwarf systems in a variety of configurations. Subsequent fragment-fragment interactions during formation can lead to inward scattering and produce a population of low semi-major axis, high-eccentricity objects. As such, the observed eccentricity of TOI-2119.01 may be an imprint of high-eccentricity migration.

\subsection{Astrometric Constraints on Additional Companions}
The high eccentricity of TOI-2119.01 may be the result of dynamical interactions with a long-period companion. We use Gaia EDR3 to probe the existence of a possible tertiary companion in the system. TOI-2119 is not listed as a likely wide binary from the study of proper motions by \cite{El-Badry2021}. Despite no clear bound companion among other Gaia sources, the precision of Gaia EDR3 allows us to probe binarity using the re-normalized unit weight error (RUWE), which is the square root of the reduced $\chi^2$ statistic that has been corrected for calibration errors, and the excess astrometric noise ($\epsilon$), a measure of the additional noise required to explain the scatter from the derived astrometric solution \citep{Lindegren2018,Lindegren2021}. \cite{Lindegren2021} note that these values are sensitive to the photocentric motions of unresolved objects. 

For orbital periods much shorter than the baseline of observations, the astrometric wobble of the primary star around the center of mass may appear as noise when adopting a single-star astrometric solution \citep[e.g.,][]{Kervella2019,Kiefer2019}. A large excess astrometric noise with a significance value, \(D > 2\), or a large RUWE has been shown to be a likely indicator of unresolved companions in recent studies \citep[e.g.,][]{Belokurov2020,Penoyre2020,Gandhi2020,Stassun2021}. TOI-2119 has an RUWE of 1.9314 and \(\epsilon=0.2644\) mas with a significance of \(D=161.3\) in Gaia EDR3. \cite{Penoyre2020} note that an RUWE \(>1.4\) could reliably be used to identify binary systems in an analysis of mock and real data of short period binaries. We use the analytical expression in Equation 17 of \cite{Penoyre2020} to calculate the projected astrometric scatter from a single-body astrometric fit, $\delta\theta$, assuming that many periods of the binary are observed, as
\scriptsize
\begin{equation}
    \delta\theta = \varpi \frac{a\lvert q-l\rvert}{\left(1+q\right)\left(1+l\right)} \sqrt{1-\frac{\sin^2 i}{2} - \frac{3+\sin^2 i\left(\cos ^2 \omega-2\right)}{4}e^2},
    \label{eq:dthetha}
\end{equation} 
\normalsize
where $a$ is the semi-major axis in au, $q=M_{2}/M_{\star}$ is the mass ratio, $l=L_{2}/L_{\star}$ is the luminosity ratio, $e$ is the eccentricity, $i$ is the inclination, $\omega$ is the argument of pericenter, and $\varpi$ is the parallax in mas. Equation \ref{eq:dthetha} should be a suitable approximation for the expected scatter due to TOI-2119.01 because Gaia EDR3's temporal baseline (34 months) is much larger than the orbital period (7.2 days). The luminosity ratio is estimated as $l\approx0.00065$ in the Gaia bandpass\footnote{\url{https://www.cosmos.esa.int/web/gaia/edr3-passbands}} using the BT-Settl models from Section \ref{sec:brighttemp}. The expected astrometric noise from all observations contained in Gaia EDR3 is $\delta\theta=0.1420$ mas. Even if we use the Gaia Observation Scheduling Tool\footnote{\url{https://gaia.esac.esa.int/gost/}} to obtain the timestamps for the Gaia EDR3 observations (79 different scans) and calculate the two-dimensional deviations at each Gaia observation following Appendix B from \cite{Penoyre2020}, we obtain a time averaged $\delta\theta=0.1406$ mas. This value is more than half of the observed $\epsilon$ and we have ignored other sources of error, such that that the observed $\epsilon$ may be completely consistent with the presence of TOI-2119.01 ($q=0.12\pm0.02$). The larger \(\epsilon\) may also be a potential indicator of an additional, long-distance companion, but we note that other scenarios, such as a close (within 0.15\arcsec), faint on-sky companion \citep[e.g.,][]{Ziegler2020,Belokurov2020} could enhance the value of \(\epsilon\). Future releases from Gaia will allow a proper motion analysis \citep[e.g.,][]{Kervella2019} of TOI-2119 to evaluate the possibility of a third body in the system.

\subsection{RV Constraints on Additional Companions}
We use {\tt thejoker} \citep{Price-Whelan2017} to perform a rejection sampling analysis on the residuals of our fit to the HPF RVs to constrain the potential mass of a third companion in the TOI-2119 system. The HPF data have a temporal baseline of 249 days and, for the rejection sampling we consider any orbits with periods $P<10000$ days. The analysis with \texttt{thejoker} uses a log-uniform prior for the period (between $8<P<10000$ days), the Beta distribution from \cite{Kipping2013a} as a prior for the eccentricity, and a uniform prior for the argument of pericenter and the orbital phase. Out of the \(>\times10^8\) (\(2^{27}\)) samples analyzed with {\tt thejoker}, a total of 56653 samples survived (\(\sim0.05\%\) acceptance rate). The masses from the surviving samples let us reject the presence of any additional low-inclination ($\sin i\sim1$) brown dwarfs ($M<11\mathrm{M_J}$) within 7.4 au of TOI-2119. Any tertiary companion, even on an inclined orbit, would need to have a period short enough for Gaia to detect binary motion \citep[$\ll 100$ years, see][]{Penoyre2020}.

\subsection{Tidal Evolution of TOI-2119}
TOI-2119.01 is on a short-period orbit with non-zero eccentricity and we may expect that tides have affected its orbit \citep[e.g.,][]{Mazeh2008,Damiani2016}. Changes to the orbit may be due to tidal torques that emerge either from the deformation of the brown dwarf by the host star or from the deformation of the host star by a brown dwarf \citep{Hut1981}. Tidal dissipation and magnetic braking from the spin-down of TOI-2119 should cause orbital decay of the brown dwarf \citep[e.g.,][]{Damiani2015}. To estimate the timescales for circularization and inspiral, we adopt the tidal model presented in Equations 1 and 2 of \cite{Jackson2008} and, similar to \cite{Persson2019}, we define the following timescales for inspiral, $\tau_a$, and circularization, $\tau_e$:
\begin{eqnarray}
    \frac{1}{\tau_{a}}&=&\frac{1}{\tau_{a,\star}}+\frac{1}{\tau_{a,\mathrm{BD}}}\label{eq:tides1}\\
    \frac{1}{\tau_{e}}&=&\frac{1}{\tau_{e,\star}}+\frac{1}{\tau_{e,\mathrm{BD}}}\label{eq:tides2}\\
    \frac{1}{\tau_{a,\star}}&=&a_{\mathrm{BD}}^{-13/2}\frac{9}{2}\sqrt{\frac{G}{M_{\star}}}\frac{R_{\star}^5 M_{\mathrm{BD}}}{Q_{\star}},\label{eq:tides3}\\
    \frac{1}{\tau_{a,\mathrm{BD}}}&=&a_{\mathrm{BD}}^{-13/2}\frac{63}{2}\sqrt{GM_{\star}^3}\frac{R_{\mathrm{BD}}^5 e_{\mathrm{BD}}^2 }{Q_{\mathrm{BD}}M_{\mathrm{BD}}}, \\
    \frac{1}{\tau_{e,\star}}&=&a_{\mathrm{BD}}^{-13/2}\frac{171}{16}\sqrt{\frac{G}{M_{\star}}}\frac{R_{\star}^5 M_{\mathrm{BD}}}{Q_{\star}},\\
    \frac{1}{\tau_{e,\mathrm{BD}}}&=&a_{\mathrm{BD}}^{-13/2}\frac{63}{4}\sqrt{GM_{\star}^3}\frac{R_{\mathrm{BD}}^5 }{Q_{\mathrm{BD}}M_{\mathrm{BD}}}\label{eq:tides6}
\end{eqnarray} 
where $\tau_{e,\star}$ and $\tau_{e,\mathrm{BD}}$ represent the contributions to the circularization timescale and $\tau_{a,\star}$ and $\tau_{a,\mathrm{BD}}$ are the contributions to the inspiral timescale due to tides raised on the star and the BD, respectively. For the tidal quality factors, we assume the brown dwarf is comparable to Jupiter and adopt a value of \(Q_{\mathrm{BD}}=10^5\) \citep[see][]{Goldreich1966,Lainey2009,Lainey2016}. This choice of $Q$ is also supported by observational evidence that brown dwarfs have \(Q_{\mathrm{BD}}>10^{4.5}\) \citep{Heller2010}. We adopt a nominal value of \(Q_{\star}=10^7\) for TOI-2119 based on the modeling of \cite{Gallet2017} and, for simplicity, assume the tidal dissipation factors remain constant. We note that $Q$ will change as the star or brown dwarf evolve \citep[e.g.,][]{Barker2009,Gallet2017}. Using the system parameters from Table \ref{tab:derivedpar}, we estimate a timescale for circularization of \(\sim56\) Gyr and a timescale for in-spiral of \(\sim221\) Gyr. In each case, the tides raised on the star by the brown dwarf (\(\tau_\star\)) dominate the timescales. The circularization timescale is \(\sim180\) Gyr even with a second order expansion to account for the moderate eccentricity of the orbit \citep[Equation 2 in][]{Adams2006}. The orbit of TOI-2119 is not evolving due to tides and should remain unperturbed for the main sequence lifetime of TOI-2119.

\begin{figure*}[!ht]
    \epsscale{1.15}
    \plotone{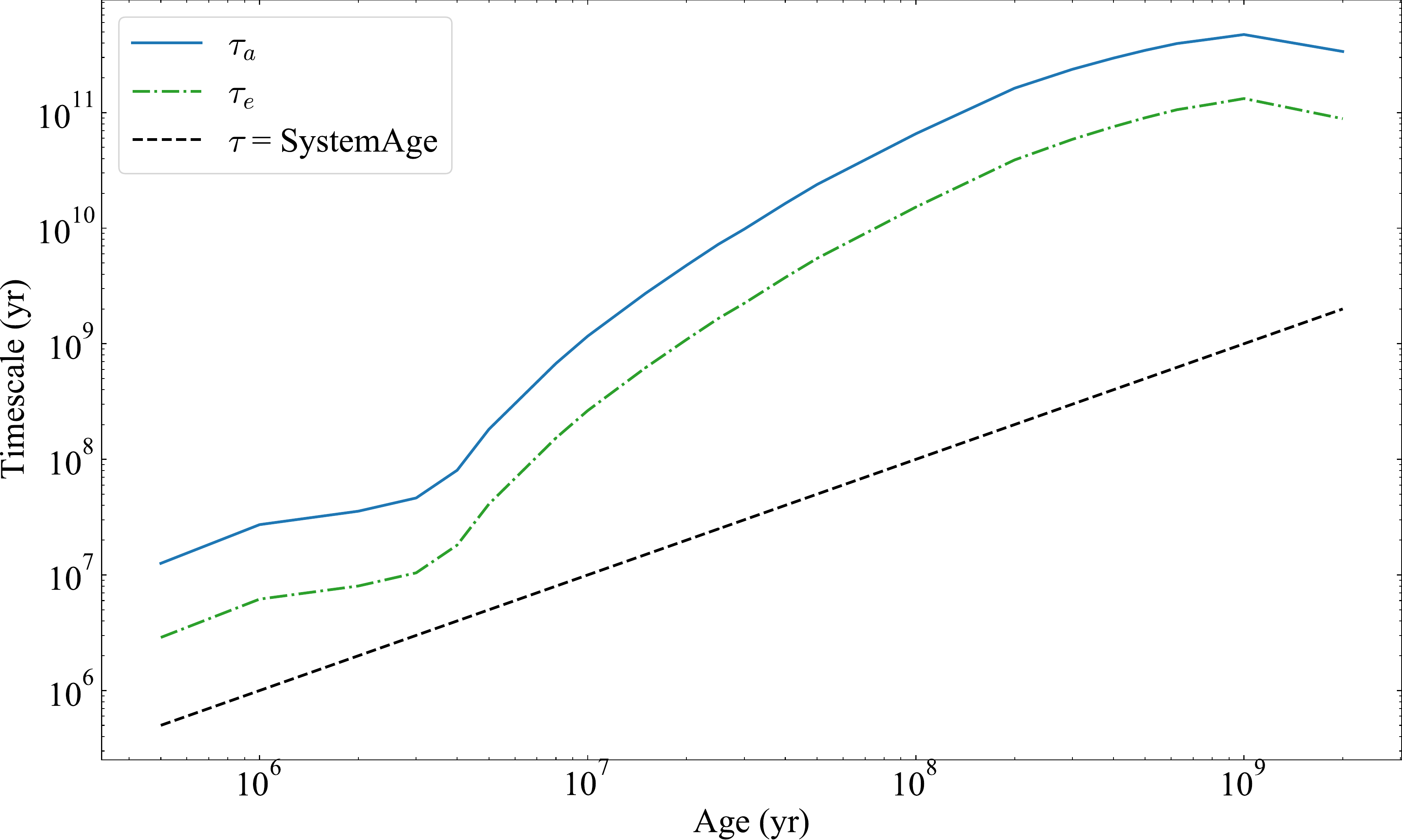}
    \caption{Above are the timescales for inspiral, $\tau_a$, and circularization, $\tau_e$, derived using Equations \ref{eq:tides1} and \ref{eq:tides2}, adopting \(Q_{\star}=10^7\) and \(Q_{\mathrm{BD}}=10^5\), and using the radii for the M dwarf and brown dwarf from evolutionary models by \cite{Baraffe2015}. The dashed line is plotted for reference and corresponds where the time scale is equal to the system's age.}
    \label{fig:2119timescales}
\end{figure*}

Equations \ref{eq:tides3}-\ref{eq:tides6} have a large dependency on the radii of each object (\(\propto R^5\)) and the timescales should change as both the host star and brown dwarf evolve. We use the evolutionary models of \cite{Baraffe2015} to obtain the radii of TOI-2119 and TOI-2119.01 as a function of time to explore the evolution of the circularization and inspiral timescales. Figure \ref{fig:2119timescales} shows that for the lifetime of TOI-2119, the ages for circularization and inspiral are orders of magnitude larger than the system age, except for ages of a few Myr when the system age is comparable to the timescale for circularization. If the observed eccentricity is primordial, TOI-2119.01 may have briefly begun circularization early in the system but then stopped as the host star contracted. The circularization timescale is only comparable to the system age for a few million years and the eccentricity of the system has probably not changed significantly due to tides. 

\subsection{Potential for Measuring the True Spin-Orbit Alignment}
The projected spin-orbit angle ($\lambda$), or the projected angle between the stellar rotation axis and the normal to the planet of the orbit, can shed light on the dynamical and formation history of a system \citep[e.g.,][]{Winn2015,Dawson2018}. Measurements of $\lambda$ for massive ($>3M_{\mathrm{J}}$), hot ($T_{e}>6000$K) planets and brown dwarfs have revealed that most of these systems are less likely to be retrograde and have lower values of $|\lambda|$ \citep[e.g.,][]{Hebrard2010,Hebrard2011,Triaud2018,Zhou2019}.  There are only a few measurements of $\lambda$ for transiting objects in the mass range \(10-80 ~\mathrm{M_J}\): HAT-P-2 b \cite{Loeillet2008}, CoRoT-3 b \citep{Triaud2009}, XO-3 b \citep{Hirano2011}, KELT-1 b \citep{Siverd2012}, WASP-18 b \citep{Albrecht2012}, and HATS-70 b \citep{Zhou2019}. These objects orbit hot stars above the Kraft break \citep{Kraft1967}, the region where stars become fully-radiative (\(T_{e}\sim6100\) K) and are observed to have low projected stellar obliquities $|\lambda|$ \citep[see][]{Zhou2019}. The lack of high $|\lambda|$ for these massive substellar companions is thought to be a result of tidal realignment, as the realignment timescale is dependent on the mass ratio \(\propto q^{-2}\) \citep[e.g.,][]{Barker2009,Dawson2014,Triaud2018}.

Unlike the current set of brown dwarfs orbiting FGK dwarfs with obliquity measurements, TOI-2119 orbits a much cooler M dwarf and any primordial misalignment should still be present because the expected timescale for spin-orbit alignment $\tau_i>10^{12}$ years \citep[e.g.,][]{Barker2009} if we adopt \(Q_{\star}=10^7\) and \(Q_{\mathrm{BD}}=10^5\). With only an upper limit of $v\sin i_\star<2$ km/s, we have no constraint on the stellar inclination and we recover a uniform distribution for $\cos i_\star$ when using the formalism of \cite{Masuda2020} to estimate the stellar inclination. If the stellar equator is well-aligned with a viewer such that $\sin i_\star=1$, we expect a rotational velocity of $v\sin i_\star=1.95\pm0.05$ km/s from our measured rotation period and stellar radius. TOI-2119 is an early M dwarf with a peak in its SED at around $\sim0.9-1$ microns which makes it possible to determine the value of $v\sin i$ using any precise optical spectrograph with a higher resolution than HPF ($R\sim55000$), such as MAROON-X \citep{Seifahrt2016}, EXPRES \citep{Jurgenson2016}, CARMENES \citep{Quirrenbach2014,Quirrenbach2018}, or NEID \citep{Schwab2016}. 

Observations with many of these precise spectrographs would also enable a detection of the Rossiter-McLaughlin effect \citep[RM effect;][]{Winn2010,Triaud2018} to measure $\lambda$. For TOI-2119, the possibility of independent measurements of $v\sin i_\star$, $\lambda$, and $P_{\mathrm{rot}}$ would allow for an estimate of the spin-orbit angle, $\psi$. $\psi$ is one of a few fundamental orbital parameters and can serve as a potential diagnostic of theories of migration \citep[e.g.,][]{Fabrycky2009}. A first order estimate for the amplitude of the RM effect is $\Delta V = 2/3 \left(R_{1}/R_\star\right)^2 v\sin i_\star \sqrt{1-b^2}$ \citep[Equation 1,][]{Triaud2018}. We estimate an amplitude of \(\sim50\) m/s for TOI-2119.01, if we assume $v\sin i_\star=1.95$ km/s, and this precision can be achieved with current spectrographs (e.g., MAROON-X, EXPRES, CARMENES, NEID). A measurement of $\psi$ in the TOI-2119 system will enable a complete dynamical characterization to inform us how this system could have formed. 

\section{Summary}\label{sec:summary}
We report the discovery of a brown dwarf ($M_{2}=67\pm2\mathrm{M_J}$ and $R_{2}=1.11\pm0.03\mathrm{R_J}$) on an eccentric (e=$0.3362 \pm 0.0005$), short period orbit ($P=7.200861\pm0.000005$ days) transiting and occulting the M dwarf TOI-2119. The rotation period of \(13.2\pm0.2\) days suggests the system probably has an age between \(0.7-5.1\) Gyrs while evolutionary models for brown dwarfs favor ages $<1$ Gyr. The difficulty in constraining the age of TOI-2119 limits our ability to use the brown dwarf companion to further constrain evolutionary models. The secondary eclipses observed by TESS reveal a temperature of $2100 \pm 80$ K for TOI-2119.01, which is consistent with measured temperatures of L0-L2 dwarfs. The high eccentricity and excess astrometric noise from Gaia EDR3 are suggestive of an additional companion in this system, but we can only exclude the existence of massive brown dwarfs on low-inclination (\(\sin i \sim1\)) orbits with our RVs. The precision of the orbital parameters of TOI-2119 will enable detailed astrometric analysis of future Gaia releases to confirm the existence of a distant, tertiary companion. The precise determination of the orbital inclination and rotation period make this system amenable to a measurement of the true spin-orbit angle with observations from high resolution spectrometers. A measurement of $\psi$ would be the first for a massive substellar companion around a cool host star and will further our understanding of the dynamical history of TOI-2119.

\section*{Acknowledgments}
We thank the anonymous referee for a thoughtful reading of the manuscript, and for useful suggestions and comments which made for a clearer manuscript. This work was supported by NASA Headquarters under the NASA Earth and Space Science Fellowship Program through grant 80NSSC18K1114 and by the Alfred P. Sloan Foundation's Minority Ph.D. Program through grant G-2016-20166039. 
The Center for Exoplanets and Habitable Worlds is supported by the Pennsylvania State University and the Eberly College of Science.

These results are based on observations obtained with the Habitable-zone Planet Finder Spectrograph on the HET. We acknowledge support from NSF grants AST 1006676, AST 1126413, AST 1310875, AST 1310885, AST 2009889, AST 2108512 and the NASA Astrobiology Institute (NNA09DA76A) in our pursuit of precision radial velocities in the NIR. We acknowledge support from the Heising-Simons Foundation via grant 2017-0494.  The Hobby-Eberly Telescope is a joint project of the University of Texas at Austin, the Pennsylvania State University, Ludwig-Maximilians-Universität München, and Georg-August Universität Gottingen. The HET is named in honor of its principal benefactors, William P. Hobby and Robert E. Eberly. The HET collaboration acknowledges the support and resources from the Texas Advanced Computing Center. 
We are grateful to the HET Resident Astronomers and Telescope Operators for their valuable assistance in gathering our HPF data. We would like to acknowledge that the HET is built on Indigenous land. Moreover, we would like to acknowledge and pay our respects to the Carrizo \& Comecrudo, Coahuiltecan, Caddo, Tonkawa, Comanche, Lipan Apache, Alabama-Coushatta, Kickapoo, Tigua Pueblo, and all the American Indian and Indigenous Peoples and communities who have been or have become a part of these lands and territories in Texas, here on Turtle Island.

Computations for this research were performed on the Pennsylvania State University's Institute for Computational and Data Sciences' Roar supercomputer, including the CyberLAMP cluster supported by NSF grant MRI-1626251.

Some of the data presented in this paper were obtained from from the Mikulski Archive for Space Telescopes (MAST) at the Space Telescope Science Institute. The specific observations analyzed can be accessed via \dataset[10.17909/t9-v3f8-w427]{https://doi.org/10.17909/t9-v3f8-w427}. Support for MAST for non-HST data is provided by the NASA Office of Space Science via grant NNX09AF08G and by other grants and contracts.
This work includes data collected by the TESS mission, which are publicly available from MAST. Funding for the TESS mission is provided by the NASA Science Mission directorate. 
This research made use of the NASA Exoplanet Archive, which is operated by Caltech, under contract with NASA under the Exoplanet Exploration Program.
This research has made use of the SIMBAD database, operated at CDS, Strasbourg, France, and NASA's Astrophysics Data System Bibliographic Services.
2MASS is a joint project of the University of Massachusetts and IPAC at Caltech, funded by NASA and the NSF.

These results are based on observations obtained with the 3\,m Shane Telescope at Lick Observatory. We acknowledge support from NSF grant AST 1910954. The authors thank the Shane telescope operators, AO operators, and laser operators for their assistance in obtaining these data.

Some of the observations in this paper made use of the NN-EXPLORE Exoplanet and Stellar Speckle Imager (NESSI). NESSI was funded by the NASA Exoplanet Exploration Program and the NASA Ames Research Center. NESSI was built at the Ames Research Center by Steve B. Howell, Nic Scott, Elliott P. Horch, and Emmett Quigley. The authors thank Mark E. Everett for assistance in obtaining these data.

These results are based on observations obtained with the Samuel Oschin Telescope 48-inch and the 60-inch Telescope at the Palomar Observatory as part of the Zwicky Transient Facility project. ZTF is supported by the National Science Foundation under Grant No. AST-2034437 and a collaboration including Caltech, IPAC, the Weizmann Institute for Science, the Oskar Klein Center at Stockholm University, the University of Maryland, Deutsches Elektronen-Synchrotron and Humboldt University, the TANGO Consortium of Taiwan, the University of Wisconsin at Milwaukee, Trinity College Dublin, Lawrence Livermore National Laboratories, and IN2P3, France. Operations are conducted by COO, IPAC, and UW.
This work makes use of data from the first public release of the WASP data \citep{Butters2010} as provided by the WASP consortium and services at the NASA Exoplanet Archive, which is operated by the California Institute of Technology, under contract with NASA under the Exoplanet Exploration Program.

This work has made use of data from the European Space Agency (ESA) mission Gaia (\url{https://www.cosmos.esa.int/gaia}), processed by the Gaia Data Processing and Analysis Consortium (DPAC, \url{https://www.cosmos.esa.int/web/gaia/dpac/consortium}). Funding for the DPAC has been provided by national institutions, in particular the institutions participating in the Gaia Multilateral Agreement.

\facilities{ASAS, Gaia, HET (HPF), PO:1.2m (ZTF), PO:1.5m (ZTF), Shane, SuperWASP, TESS, WIYN (NESSI)} 
\software{
\texttt{allesfitter} \citep{Guenther2019},
\texttt{astroquery} \citep{Ginsburg2019},
\texttt{astropy} \citep{AstropyCollaboration2018},
\texttt{barycorrpy} \citep{Kanodia2018}, 
\texttt{dynesty} \citep{Speagle2020},
\texttt{ellc} \citep{Maxted2016},
\texttt{EXOFASTv2} \citep{Eastman2019},
\texttt{HPF-SERVAL},
\texttt{HPF-SpecMatch},
\texttt{juliet} \citep{Espinoza2019},
\texttt{matplotlib} \citep{hunter2007},
\texttt{numpy} \citep{vanderwalt2011},
\texttt{pandas} \citep{McKinney2010},
\texttt{scipy} \citep{Virtanen2020},
\texttt{telfit} \citep{Gullikson2014},
\texttt{thejoker} \citep{Price-Whelan2017}
}

\bibliography{combined}

\end{document}